%% file: icmCompressibilityfinal.tex
\documentclass[a4paper,10pt,twoside]{revtex4}
\usepackage[utf8]{inputenc}%
\usepackage[left=1in,right=1in,top=0.8in,bottom=0.8in,includefoot,includehead,headheight=13.6pt]{geometry}
\usepackage{amsmath}
\usepackage{amssymb}%
\usepackage{graphicx}
\usepackage{comment}
\usepackage{color} 

\include{newcommands}
\include{myNewCommands}

\graphicspath{{images/}}

\begin{document}

\title{A minimal model for acoustic forces on Brownian particles}
\author{F. Balboa Usabiaga}
\email[]{florencio.balboa@uam.es}
\author{R. Delgado-Buscalioni}
\email[]{rafael.delgado@uam.es}
\affiliation{Departamento de Física Teórica de la Materia Condensada and IFIMAC. 
Universidad Autónoma de Madrid,Campus de Cantoblanco, Madrid 28049, Spain}
\begin{abstract}
We present a generalization of the {\em inertial coupling}
(IC) [Usabiaga  {\em et  al.}  {\em J.   Comp. Phys.}  2013]  which permits the resolution
of radiation forces on small particles with arbitrary acoustic contrast factor.   The IC
method  is  based  on   a  Eulerian-Lagrangian approach:
particles move in continuum space while the fluid equations are solved
in  a regular  mesh (here we use the finite volume method).  Thermal
fluctuations in  the fluid stress,  important below the  micron scale,
are  also  taken  into  account following  the  Landau-Lifshitz  fluid
description.  Each particle is  described by a minimal cost resolution
which  consists  on  a  single  small  kernel  (bell-shaped  function)
concomitant to the particle. The  main role of the particle kernel is
to interpolate fluid properties and spread particle forces.  
Here, we  extend the  kernel  functionality to  allow for  an
arbitrary particle compressibility.  
The particle-fluid force is obtained from an imposed  ``no-slip'' constraint  
which  enforces similar particle and kernel fluid velocities. This coupling  is {\em instantaneous}  
and permits  to capture the fast, non-linear effects underlying the radiation forces on particles.   
Acoustic forces arise either because an
excess in particle compressibility (monopolar term) or in mass  (dipolar   contribution)  over   the   fluid  values.
Comparison  with theoretical  expressions  show   that  the  present
generalization of the IC  method correctly reproduces both contributions.  Due
to  its  low  computational   cost,  the  present  method  allows  for
simulations with many [$O(10^4)$] particles using a standard
Graphical Processor Unit (GPU).
\end{abstract}

\maketitle

\section{Introduction}
Sound waves  in the ultrasonic  frequency range $\omega>\mathrm{kHz}$,
are used for an amazing list of applications such as object detection,
testing flaws  in materials, medical  imaging, cleaning, therapeutic al
purposes,  tumor   destruction,  and  even  as   weapon.   A  related
phenomena,  cavitation,   uses  powerful   kHz  waves  to   produce  a
significant  temperature  and  pressure  increase in  the  liquid  and
locally  boost chemical  reactions.   At larger  MHz frequencies,  the
sound wavelength  in a  typical liquid is  in the millimeter  range and
thus suited for lab-on-a-chip technologies \cite{Settnes2012}. MHz  sound
interacts and impinges  forces to micron-size particles due  to a nice
example of non-linear correlation  between the oscillating density and
velocity  fields  \cite{Settnes2012}.   Such  force, known  as  acoustic
radiation  force  \cite{Settnes2012},  was theoretically  predicted  for
rigid objects in a fluid by King \cite{King1934} in 1934 and two decades
later   extended   to   compressible   particles   by   Yoshioka   and
Kawashima\cite{Yosioka1955}.      In     the     sixties,     Gor'kov
\cite{Gorkov1962}  published   an  elegant  approach   in  the  soviet
literature,  showing   that  in  the  inviscid   limit  (large  enough
frequencies) the radiation force for standing waves can be derived from  the gradient of
an effective  potential energy $\bF_{ac}  = -\nabla U_{ac}$;  a result
that    has   been   quite    useful   for    subsequent   engineering
applications.   The  (sometimes   called \cite{Wang2011})   Gor'kov   potential, scales with the  particle volume $\vol$
and has contributions from the time-averaged pressure  $p_{in}$ and
velocity $v_{in}$ of the incoming wave,
\begin{equation}
\label{eq:acousticPotential}
  U_{ac}= -\frac{\vol}{2} \left[\kappa_e \langle p_{in}^2\rangle + \frac{3 m_e}{2m_p+\rho_0 \vol} \rho_0 \langle v_{in}^2\rangle\right],
\end{equation}
where $\rho_0$ is the fluid density and
$\langle x \rangle =  (1/\tau) \int_0^\tau  x(t) dt$ 
and $\tau=2\pi/\omega$ is the oscillation period.
These two contributions to the acoustic potential (\ref{eq:acousticPotential})
are proportional to particle excess-quantities 
relative to the fluid  values. 
In particular, $m_e =  m_p -\rho_0 \vol$ 
denotes the excess of particle mass ($m_p$) 
over the mass of fluid it displaces $\rho_0 \vol$
and $\kappa_e  = \kappa_p  -\kappa_f$ 
is the excess in particle compressibility ($\kappa = (1/\rho)
\partial \rho/\partial p$) relative to the fluid.
Despite their  relevance, the early papers on  acoustic radiation were
rather scarce in explanations  and recent theoretical works revisiting
this  phenomenon  have been  most  welcome  (see \cite{Settnes2012}  and
citations  thereby).   Bruus   \cite{Settnes2012}  used  a  perturbation
expansion in  the (small)  wave amplitude to  show that both  terms in
Eq. \ref{eq:acousticPotential}
are in  fact related to the monopole and dipolar moments
of the flow potential, which are uncoupled
in  linear acoustics.  He  also extended  the
analysis  to  the  viscid  regime (smaller  frequencies)  generalizing
previous  studied  by Doinikov  \cite{Doinikov1994,Doinikov1997}  and  others  (see
\cite{Settnes2012}).

The first application of ultrasound forces were carried out 
in the eighties by Maluta
{\em et al.} \cite{Dion1982}.
They used standing  waves to trap and orient  wood pulp fibers diluted
in water  into the equidistant pressure  planes. The idea  was used by
the paper industry  to measure the fiber size.   Recently the usage of
ultrasound  for  manipulation  of  small  objects  is  flourishing  and
offering  many promising applications  for material  science, biology,
physics,  chemistry and  nanotechnology.  An  excellent review  of the
current  state-of-art can  be found  in the  monographic issue  on the
journal {\em Lab on a chip} [Volume 12, (2012)] and also 
in the review of Ref. \cite{Friend2011} which focuses on applications,
cavitation and more exotic phenomena. Trapping extremely small
objects (reaching submicron-sizes) using  ultrasound, in what has been
called  ``acoustic  tweezers''  \cite{Ding2012}, is explored  by
several groups \cite{Oberti2007,Ding2012}
 and used for  many different purposes, such as
to  move and capture  colloids \cite{Oberti2007}  or even  individual living
cells  without  even  damaging  them\cite{Haake2005a,Ding2012}.  
Quoting  T.J.Huang:  ``acoustic tweezers are much smaller than  optical tweezers and use 500,000 times less energy.''\cite{Ding2012}

Despite the increase in  theoretical and experimental works, there are
not too many numerical simulations on ultrasound-particle interaction.
Its cause might be  the inherent difficulties  this phenomenon poses
to  numerical calculations. The  acoustic force  arises  as a
non-linear  coupling between two  fast-oscillating signals
and only manifests  after averaging  over  many oscillations.   
This means a 
tight connection between the fastest hydrodynamic mode (sound) and the
much slower viscous motion of the particle, at 
a limiting velocity dictated by the viscous drag.  
The situation, from the numerical standpoint,
is even worse if one is  interested in studying the dispersion of many
small colloids around the loci of the minima of the Gor'kov potential,
because  dispersion is  a diffusion-driven  process and  requires much
longer time scales.  Colloidal dispersion around the accumulation loci
is certainly important  and a nuisance for many  applications.  It was
first studied by Higashitani {\em et al} \cite{Higashitani1981}, 
who worked with the hypothesis that
the particles follow a Boltzmann distribution based on
the acoustic potential  energy. Simulation of 
a swarm of particles diffusing under acoustic radiation
involve solving an intertwined  set of  mechanisms acting
over time-scales spanning over many  decades.   As a  typical
example,  in  a  liquid,   sound  crosses  a  micron-size  colloid  in
$R/c_F\sim 10^{-9}$ seconds, while the colloid diffuses its own radius
in  $R^2/D  \sim  10^0$  seconds. Such wide  dynamic range 
is certainly impossible to tackle for any numerical method 
involving a detailed resolution of each particle surface.

An important task for numerical studies in the
realm of  acoustic force applications 
is the determination of the pressure pattern in resonant cavities
 \cite{Skafte-Pedersen2008,Dual2012}.
The main objective of these calculations,
which solve the Helmholtz wave equation 
(but do not involve any particle) is to 
forecast the pressure nodes inside the chamber, where 
colloidal coagulation is expected to occur.
Using a {\em one-way-coupling} approach \cite{Maxey1983},
it is also possible to get some insight on the particle trajectories,
by directly applying the theoretical acoustic forces together with the 
(self-particle) viscous drag \cite{Muller2012}. 
This leads however to uncontrolled approximations \cite{Skafte-Pedersen2008}
which neglect significant non-linear 
effects such as the hydrodynamic particle-particle interactions 
and the effect of multiple particle scattering on the wave pattern 
\cite{Feuillade1996}.

Another group of numerical studies explicitly calculate the acoustic force
on  objects although,  to the  best of  our knowledge,  have been  so far
restricted  to  single  two-dimensional  spheres (or,  more  precisely
axially projected  "cylinders") \cite{Wang2009}.  These  works were
based on finite element or  finite volume discretizations of the fluid
and  the  immersed object,  with  explicit  resolution  of its  surface
(no-slip and impenetrability conditions).  The effect of viscous
loss has been  studied in  a recent  work \cite{Wang2011}.
There are  also some
calculations using  Lattice Boltzmann solvers  \cite{Barrios2008} also involving
single  2D cylinders and ideal fluid.
It  has to  be mentioned  that all these works
considered rigid  particles.  In fact, implementing  a finite particle
compressibility   is    not   straightforward   for this   type   of
surface-resolved  approach  as it  would  demand implementing  elastic
properties to the solid and couple  it to the dynamics of the particle
interior\cite{Hasegawa1979}.  Another
downside of  fully fledged resolution is the  large computational cost
per  particle which limits  feasible simulations  to few  particles at
most.

In  this work  we propose  a quite  different modeling  route  for the
particle   dynamics.    First,   our    method   is   based   on   the
Eulerian-Lagrangian  approach \cite{Duenweg2009,Peskin2002}, meaning
that particles are  not constructed with or restricted  to the ``fluid
mesh''  but   move  freely  in  the  continuum   space.   This  avoids
complicated  triangulation and remeshing  around the  particle and
permits   solve  the  fluid   equations  (we   consider  Navier-Stokes
Fluctuating Hydrodynamics) in a simple regular lattice of fluid cells,
using a finite volume scheme \cite{Usabiaga2012}.   Second, particles are
described with  a minimal-resolution  model involving a  single kernel
function per  particle, which just contains $3^3$ fluid-cells  in 3D.
The  particle   kernel,  originally   designed  by  Peskin   and  Roma
\cite{Roma1999}  for  the  Immersed  Boundary  (IB) method,  is  used  to
interpolate local  fluid properties and to spread  the particle forces
to the surrounding fluid.  The  third important issue, and in fact the
novelty of what we refer to  as  ``inertial
coupling'' (IC)  method \cite{Usabiaga2013} resides  in imposing an
{\em instantaneous} ``no-slip''  constraint (the particle velocity equals
the  interpolated  fluid velocity)  to  couple  the  dynamics of the particle
and  the fluid.   Such coupling  is  {\em instantaneous}  and, as   shown  in  our  previous  work \cite{Usabiaga2013}
it captures the  fast ultrasound-particle  interaction. 
Here we further explore this line of {\em minimally-resolved particle modeling} which is based on the idea that
the  particle kernel  (originally designed  
for interpolation   purposes  \cite{Peskin2002}) can be used to embed  all the
relevant physical properties of the particle, such as its  hydrodynamic radius $R_H$  \cite{Duenweg2009}, its
volume $\vol$, and mass ($m_p=m_e+\rho_\vol$).   A
characteristic feature  of this minimal model 
(which proves to be beneficial for the present work) is the absence 
of density boundary conditions to ensure the particle impenetrability across its surface.
In fact, in the present model the ``particle'' has not a well-defined  surface
and the fluid density field is not zero inside the particle domain. For this   reason, after  Dunweg  and
Ladd \cite{Duenweg2009}, this model is sometimes called ``blob'' model.  Imposing   a  pressure  force  $\bF$   to  a  surface-less
``particle'' is however not a problem provided it is contained in a well defined volume $\vol$.  
Thanks to  Gauss-Ostrogradsky integral theorem, we can
convert  the traction done by pressure (tensor) ${\bf P}$
over the particle surface  $S$ to an integral over its volume $\vol$
\begin{equation}
  \bF = -\oint_S {\bf P} \cdot \bn \,\dd r^2 = -\int_{\vol} \nabla \cdot {\bf P} \dd r^3 = -\vol  \int \Theta({\bf q} -\br) \nabla \cdot {\bf P} \dd r^3.
\end{equation}
where ${\bf q}$ is the position of the particle center. The second equality is 
indeed exact for the hard-kernel of a rigid particle $\Theta(\bq-\br)$ which differs from zero only inside the particle, where
\begin{equation}
\label{hker}
\Theta({\bf q} -\br)= 1/\vol \;\;\mbox{for} |\br-\bq| < a.
\end{equation}
The ``blob'' approach consist on deploying instead a soft-kernel
(bell-shape, everywhere  derivable). Slightly different version of this idea is used in all
Eulerian-Langrangian and fully Lagrangian (meshfree)
particle methods \cite{Lomholt2003,Tatsumi2012, Vazquez-Quesada2012}. 

The IC  method for  particle hydrodynamics was  presented in  a recent
work \cite{Usabiaga2013} and  subsequently extended to incompressible
flow \cite{Usabiaga2012b}.   Capturing ultrasound forces was  one of the
relevant tests performed  \cite{Usabiaga2013} to check the viability
of its {\em instantaneous  coupling}. However, as stated, the original
blob model does  not impose any constraint on  the fluid density field
and, not unexpectedly, the resulting  acoustic forces were found to be
fully compatible with particles with the same compressibility than the
fluid,   i.e.   to   $\kappa_e=\kappa_p-\kappa_f=0$.    Inspection  of
Eq. (\ref{eq:acousticPotential}) indicates  that our neutrally buoyant ``blobs''
$m_e=0$  did not experienced  any irradiation  force.  In  the present
work we focus on the acoustic  force problem and extend the blob model
to allow  for a particle  compressibility, different from that  of the
fluid.  This  is part  of a  research line with  two main  targets: to
extend the kernel functionality  by assigning more physical properties
to  it  and  more  generally,  to highlight  that  a  carefully  built
minimally-resolved model can achieve considerable accuracy and capture
realistic physics over a broader range of time and length scales.

We start by presenting the essential kernel properties in Section \ref{sec:particleModel}
and focus in how to implement the particle compressibility in Sec. \ref{compress}.
The dynamics of the particle and fluid coupled equations of motion is described in Sec. \ref{ICM},
where it is shown that the model preserves the local momentum and also 
the energy in the ideal fluid limit. It is then shown that equilibrium fluctuations
(of velocity and particle density) are consistent with the thermodynamic prescriptions.
Acoustic forces are briefly reviewed in Sec. \ref{acousticforces}. Simulations,
presented in Sec. \ref{simulations}, are shown to agree with the theoretical 
monopolar and dipolar primary forces. A study of the dispersion of a small colloid
under a standing wave is also presented. Concluding remarks are finally given in Sec. \ref{conclusions}.

\section{Particle model: kernel properties}
\label{sec:particleModel}

One of the most important  issues in the blob-particle approach is the
construction  of the  particle  kernel $\ker(\br-\bq)$.  
From  the  standpoint of  the hybrid Eulerian-Lagrangian methodology, the role of the kernel is to
act as the ``glue'' between both descriptions. As carefully explained in
previous  works \cite{Usabiaga2012b,Atzberger2011},  the kernel
provides  the two translating operations: the {\em averaging} operator transfers information from the 
Eulerian representation of the fluid to the Lagrangian
representation of the particles
$\J :\mathcal{E}  \rightarrow   \mathcal{L}$   while
the the {\em spreading} operator $\S :\mathcal{L} \rightarrow \mathcal{E}$,  translates 
``Lagrangian'' forces into ``Eulerian'' force density fields.  
 
In the continuum formulation, these two operations are defined as
\begin{eqnarray}
\label{Jcont}
\J(\bq) \bv(\br) &=& \int \ker(\bq-\br)\bv(\br) \;d^3 r,\\
\label{Scont}
\S(\bq) \bF(\bq) &=& \ker(\bq-\br) \bF(\bq),
\end{eqnarray}
so it is clear that $\S$ has units of inverse volume.
As noted in \cite{Usabiaga2012b}, using the same kernel to spread and interpolate,
brings about an important mathematical property which is crucial to maintain energy conservation and
the fluctuation dissipation balance: $\J$ and $\S$ are {\em adjoint},
\begin{equation}
\label{adj}
 \J\bv \cdot \bu = \int \bv\cdot \S\bu \,d^3r = \int \ker(\bq-\br) \bv\cdot \bu d^3r
\end{equation}

The Eulerian fluid description is solved in a discrete mesh,
which for practical purposes is regular, $\br_{\vec k} = h {\vec k}$.
Therefore, in practice, one needs to work with the discrete version of Eqs. \ref{Jcont},
\begin{eqnarray}
\label{Jdis}
\J(\bq) \bv(\br) &=& \sum_i h^3 \ker(\bq-\br_i)\bv(\br_i) \\
\label{Sdis}
\S(\bq-\br_i) \bF(\bq) &=& \ker(\bq-\br_i)\bF(\bq) 
\end{eqnarray}
where $h^3$ is the volume of the hydrodynamic cell. Discreteness brings about restrictions in the kernel shape. First,
the operation $\J$ becomes a discrete average which should at least have linear consistency:
i.e. for {\em any} Lagrangian position $\bq$,
\begin{eqnarray}
\label{k1}
  \sum_i h^3 \ker(\br_i-\bq) &=& 1\\
\label{k2}
  \sum_i h^3 (\br_i-\bq) \ker(\br_i-\bq) &=& 0.
\end{eqnarray}
This ensures that any linear field $f(r)=a+br$ is exactly interpolated, $f(q)= \sum_i h^3 f(r_i) \ker(r_i-q)$.

\subsection{Kernel volume}

In the blob-model approach, the particle kernel is not only sought as mathematical object,
but also a tool to provide {\em physical} meaning to the particle model.
This idea is clearly illustrated with the kernel volume, which in fact,
introduces the third condition in the kernel construction.
Note that the norm of the hard-kernel (\ref{hker}) trivially yields the inverse volume
of the domain, $\int \Theta({\bf q} -\br)^2 d^3r = 1/\vol$. 
Similarly, in the discrete Eulerian mesh, the norm of the kernel,
\begin{equation}
\label{kvol}
  \J\S= \sum_i h^3 \ker(\br_i-\bq)^2 = 1/\vol,
\end{equation}
should be independent on the Lagrangian position $\bq$. 
Although for different reasons, this condition (\ref{kvol}) was first formulated by Peskin \cite{Peskin2002}
in his Immersed Boundary (IB) method. In fact, conditions (\ref{k1}),(\ref{k2}) and (\ref{kvol}) determine
the 3-point kernel introduced by Roma and Peskin \cite{Roma1999}, whose norm,  in 1D, is $(1/2)h$.
For 3D, the standard tensor product construction, 
$\ker(\br) =\ker(x)\ker(y)\ker(z)$, which trivially yields, $\vol =8h^3$.
Thus, the ``blob'' volume cannot be arbitrary changed, being a property of the kernel.

\subsection{Hydrodynamic radius}

The kernel provides all the relevant physical dimensions of the ``blob''.
In previous works \cite{Usabiaga2012b,Usabiaga2013} we measured
its hydrodynamic  radius [$R_H=(0.91 \pm 0.01) \,h$, where 
the error bar comes from the variation of $R_H$ over the mesh]
from the ratio 
between a drag force $F_d$ and the 
resulting fluid terminal velocity $v_0$,
at small Reynolds number, $R_H=F_d/(6\pi\eta v_{0})$. Fitting the perturbative flow created around the blob 
to the Stokes profile gave a similar value of $R_H$ \cite{Usabiaga2013}.
The size of the perturbative  vorticity field  created by the particle is
related to its hydrodynamic radius and can be also estimated from 
its effective Fax\'en radius. The perturbative velocity field $\bv(r)$ 
created by an immersed sphere at $\br=\bq$ can be expanded as $\bv(\br)=\bv(\bq) + (a^2/6) \nabla^2\bv(\bq)+...$ \cite{Kim-Karrila}.
The Fax\'en term is proportional to the squared particle radius $a^2$. Taylor expanding $\bv(r)$ around $\bq$, 
\begin{equation}
  \bv(\br)= \bv(\bq) +\nabla \bv(\bq) (\br-\bq) +\frac{1}{2}  (\br-\bq)^{T} \cdot \nabla \nabla \bv(\bq) \cdot (\br-\bq) + ...
\end{equation}
and applying the average operator $\J$ yields,
\begin{equation}
\label{faxen}
  \J \bv(\br)= \bv(\bq) + \frac{1}{2}  \nabla^2 \bv(\bq) \J\left[(r-q)^2\right] + O(\J\left[(r-q)^4\right]),
\end{equation}
which informs  about the  effective Fax\'en radius  of our  blob model
\cite{Usabiaga2012b}: $R_F^2= 3 \J\left[(r-q)^2\right]$. For the 3-pt kernel this gives
$R_F=0.945\,h$ with a small variation of about $5\%$ over the mesh.

\section{Blob compressibility}
\label{compress}

In this work the idea of adding physical properties to the blob, via the kernel, is extended
to provide a finite blob compressibility. To that end we use the kernel to include
a local particle contribution to the pressure equation of state.
The idea is thus quite general and independent of the type of
particle-fluid coupling used and of the equations of motion (presented in Sec. \ref{ICM}),
although here we solve the isothermal compressible Navier-Stokes equations.
The pressure of the fluid phase is barotropic $p=p(\rho)$ and
we consider $p(\rho)=p_0+c_f^2 \rho$, with constant speed of sound $c_f$. 
To take into account the effect of a compressible particle 
in the fluid we propose a modification to the pressure field based on 
the following functional,
\begin{equation} 
\label{eq:pressure}
\pi(\rho,\bq) = p(\rho) + \S(\bq) \Omega(\rho;\bq).
\end{equation}
The extra  particle contribution $\S(\bq- {\bf r}) \Omega(\rho;\bq)$
only affects locally within each  particle domain. Recall that $\S$ has
units of  inverse volume, so $\Omega$ has  dimensions of energy.
In  fact, the field $\S\Omega$ can be related to the chemical potential 
created by particle-fluid interactions  \cite{pep-rafa}  (see Sec.\ref{conclusions}).
It determines  the energetic cost  for fluid  entertainment into  the kernel
domain. A simple, yet  efficient, implementation of $\Omega$ consists
on assuming that the particle contribution to the pressure is a linear
function of the averaged local density,
\begin{equation}
\label{omega}
\Omega(\J\rho) =  \epf \vol \pare{\J\rho - \rho_{0}}  
\end{equation}
where $\rho_{0}$ is the fluid equilibrium density and 
the auxiliary parameter $\epf$ is the 
particle-fluid interaction energy per unit of fluid mass \cite{pep-rafa}.
Note that $\Omega$ depends on $\bq$ through the average operator $\J=\J(\bq)$.
A variation in $\Omega$ corresponds to a work done by the fluid to compress 
the particle  domain, or  more precisely to increase  the fluid  density inside the fixed  volume $\vol$ 
(which surrounds the particle  and moves  along with  it).  
The  particle   mass  can   assigned   to  be
$m_p=m_e+\rho_0 \vol$, where $m_e$ is the excess of particle mass over
the  mass of  fluid it displaces {\em  in  equilibrium} ($\Omega =0$).
Thus, in Eq.  (\ref{omega}) we choose $\Omega$ to be
proportional to the mass of fluid $\vol  (\J\rho-\rho_0)$ 
that have entered into the  kernel domain. The resulting fluid work is positive if
the particle is compressed and  viceversa.  We will come back to this issue
in next section where the equation of motion of the blob is derived.

One can now evaluate the compressibility $\kappa(\br)$ and the speed
of sound $c(\br)$ of the fluid, which are scalar fields. 
To that end we evaluate the pressure variation $\delta \pi(\br)$ 
\begin{equation}
  \delta \pi(\br) = \int \frac{\delta \pi(\br)}{\delta \rho(\br')} \delta \rho(\br') d^3r
\end{equation}
where the functional derivative $\delta \pi(\br)/\delta \rho(\br')$ 
provides change of the pressure field at $\br$ (per unit volume) 
due to a density perturbation $\delta \rho(\br')$.
The total pressure functional can be written as
\begin{equation}
  \pi(\br) = \int p[\rho(\br')] \delta(\br'-\br) d^3 r' + \epf \vol \ker(\bq-\br) \int \ker(\bq-\br') (\rho(\br')-\rho_0) d^3 r',
\end{equation}
whose functional derivative is given by 
\eqn
\frac{\delta \pi(\br)}{\delta \rho(\br')} = c_f^2 \delta(\br-\br') + \epf \vol \ker(\bq-\br)\ker(\bq-\br').
\eqnend 
where,  $c_f$ is constant for the fluid equation of state used hereby 
(in general $c_f^2(\br)=\partial p(\rho(\br))/\partial \rho$ is a density dependent field).

In terms of the spreading and average operators, the pressure first variation is then
\begin{equation}
\label{dp}
  \delta \pi(\br) = c_f^2 \delta \rho(\br) + \epf \vol \S \J(\delta \rho).
\end{equation}

A sound velocity field $c(\br)$ 
can be defined as
\begin{eqnarray}
  c^2(\br) &=& \int \frac{\delta \pi(\br)}{\delta \rho(\br')} d^3r' = c_f^2 + \epf\vol \S(\bq-\br)
\end{eqnarray}

Averaging in (\ref{dp}) gives  the overall variation of pressure inside the kernel
which, for constant fluid sound velocity $c_f$ is equal to,
\begin{equation}
\label{jdp}
  \J[\delta \pi] = \left(c_f^2+\epf \right) \J[\delta \rho],
\end{equation}
where we have used $\J\S=\vol^{-1}$. Equation (\ref{jdp}) can be understood as the blob equation of state, 
which justify our identification of $c_p$ with the speed of sound inside the particle. It is given by,
\begin{equation}
\label{part_svel}
c_p = \sqrt{c_f^2 +\epf}.
\end{equation}

The input parameter $\epf$ can be then either positive or negative (with the obvious condition $c_p \ge 0$).
For instance, taking $\epf \simeq -c_f^2$ permits to simulate very compressible particles (gas bubbles).
Equivalently, one can introduce $\kappa_p=\kappa_f+\kappa_e$ 
where $\kappa_p \equiv 1/(\rho_0 c_p^2)$ and $\kappa_f= 1/(\rho_0 c_f^2)$
provide the particle and fluid compressibility, respectively. Then
using (\ref{part_svel}), the ``excess particle compressibility'' is just 
\begin{equation}
\label{ke}
  \kappa_e = -\frac{\epf}{c_p^2} \kappa_f.
\end{equation}
It is noted that the term related to the particle compressibility in the
ultrasound potential of Eq. (\ref{eq:acousticPotential}) is proportional to $\kappa_e$
but either $\epf$ or $\kappa_e$ can be used as input parameters of the model.

From Eq. (\ref{dp}) one can also infer a bulk modulus operator which applied to
any density perturbation field $\delta \rho(\br)$ provides the 
resulting variation in the pressure field $\delta \pi(\br)= \mathcal{B} \delta \rho(\br)/\rho_0$,
\begin{equation}
\label{km1}
 \rho_0^{-1} \mathcal{B} \equiv c_f^2 {\bf 1} + \epf \vol \S \J
\end{equation}
Its inverse ${\bs \kappa}=\mathcal{B}^{-1}$ is the compressibility operator,
which applied to some pressure field $\delta \hat p(\br) $ provides the resulting density perturbation
$\delta \hat \rho(\br) = \rho_0 {\bs \kappa}[\delta \hat p(\br) ]$.
To invert (\ref{km1}) one can use the same formal Taylor expansion used in appendix A of Ref. \cite{Usabiaga2012b} 
and get,
\eqn
\label{kappao}
{\bs \kappa} = \kappa_f + \kappa_e \vol \S \J.
\eqnend

\section{Inertial coupling method}
\label{ICM}

\subsection{Coupling}

In this section  we present the essence of  the Inertial Coupling (IC)
method   \cite{Usabiaga2013,Usabiaga2012b},   developed   to   capture
inertial effects in simulation of colloids and other microparticles in
compressible  or incompressible  flows. The IC method  uses ingredients 
of the Immersed  Boundary (IB)  method \cite{Peskin2002}, 
and in particular those related to how to ``hide'' the discrete mesh to the kernels.
Here  however,  each  kernel is not a surface-marker, but represents a single particle 
whose  dynamics should   be   infered   from   some  suitable   {\em   coarse-grained}
representation  of  the  constraints it imposes to the fluid  velocity. 
In particular,  the fluid  velocity  at the  boundary of  a
spherical particle with a non-slip surface should satisfy,
\begin{equation}
  \bv(\br) = \bu + {\bs \omega} \times \left(\br-\bq\right) \;\;\mbox{for} |\br -\bq| \leq a,
\end{equation}
where $a$ is the particle radius, $\bu$ its translation velocity,
${\bs \omega}$ its angular velocity and $\bq$ its center position.
Applying the average operator in the previous equation and noting that
$\J(\bq)[\br-\bq]=0$ one gets a {\em coarse-grained} representation of the no-slip constraint,
\begin{equation}
\label{noslip}
  \J(\bq) \bv(\br) = \bu.
\end{equation}
which is the one implemented in the present method.  
The constraint (\ref{noslip}) does not resolve the effect of particle 
rotation and rigidity (no strain) on the surrounding fluid
(see Refs. \cite{Yeo2010} for generalizations). 
The no-slip constraint is non-dissipative, so it conserves the  energy  of the  fluid-particle  system 
in reversible  processes (i.e. in  the inviscid limit)  \cite{Usabiaga2012b}. 
The no-slip constraint (\ref{noslip}) can be  generalized to  allow for  partial slip  (see
Appendix B of \cite{Usabiaga2012b}) which introduces a finite relaxation
time ($m/\xi \sim \mu \mathrm{s}$)  for the equilibration of the
particle and local fluid velocities \cite{Bedeaux1974}.  Partial-slip
dissipates energy and requires adding an extra random force 
to represent the transmission of momentum (tangential to the particle surface)
through fluid-particle molecular collisions and to 
guarantee the fluctuation-dissipation balance.  By contrast, the  no-slip  constraint idealizes 
{\em instantaneous} fluid-particle interactions which, in practice captures the  extremely fast
forces involved in the acoustic time  scale ($a/c  \sim  10^3
\mathrm{ps}$), which are actually not far from molecular forces decorrelation times \cite{Kinsler_book}.

\subsection{Dynamics}

In this section we present the equations of motion for the fluid and 
a single particle (the generalization to $N$ particles is straightforward).
These equations were discussed in previous works \cite{Usabiaga2013,Usabiaga2012b}
and the novelty here is the addition of the particle compressibility contribution 
in the pressure field  $\pi=\pi(\rho,\bq)$, whose details were discussed
in Sec. \ref{compress}. The fluid and particle dynamics
are specified by the conservation of fluid mass and momentum
[Eqs. (\ref{eq:consvDens}) and (\ref{eq:consvMom})], the particle momentum
Eq. (\ref{eq:partVel}) and the (no-slip) fluid-particle coupling  (\ref{eq:noSlip}),
\eqn
\label{eq:consvDens}
\ps{t} \rho + \bna \cdot \bg &=& 0 \\
\label{eq:consvMom}
\ps{t} \bg + \bna \cdot (\bg\bv) &=& -\bna\cdot\bP - \S \bl \\
\label{eq:partVel}
m_e \dot{\bu} &=& \bF(\bq,t) + \bl \\
\label{eq:noSlip}
\mbox{s.t. } \bu &=& \J \bv.
\eqnend
The total stress tensor is now given by,
\begin{equation}
\label{totp}
\bP=\pi {\bf 1} - \bs{\sigma} = p(\rho) {\bf 1} + \S(\bq-\br) \Omega - \bs{\sigma},
\end{equation}
where the particle-fluid interaction energy $\Omega$ is given by Eq. (\ref{omega}).
We consider a Newtonian fluid, with constant shear and bulk viscosities $\eta$ and $\zeta$
and this allows us to write the divergence of the viscous terms in the
standard Laplacian form,
\eqn
\bna \cdot \bs{\sigma} = \eta \bna^2 \bv + \pare{\zeta + \fr{\eta}{3}} \bna (\bna \cdot \bv) + \bna \cdot \bs{\Sigma},
\eqnend
The stochastic components of the
stress tensor are collected in $\bs{\Sigma}$ \cite{Landau1987,Fabritiis2007,Donev2010b,Usabiaga2012,Usabiaga2013},
being given by 
\eqn
\bs{\Sigma} = \sqrt{2\eta\kt} \wtil{\mc{W}} + \pare{\sqrt{\fr{\zeta\kt}{3}}-\fr{1}{3}\sqrt{2\eta\kt}}
\mbox{Tr} \pare{\wtil{\mc{W}}} \II.
\eqnend
Where the symmetric tensor $\wtil{\mc{W}}=(\mc{W}+\mc{W}^T)/2$ 
is defined by the covariance of a random Gaussian tensor $\mc{W}$
delta-correlated in time and space,
\eqn
\ang{\mc{W}_{ij}(\br,t)\mc{W}_{kl}(\br',t')} = \delta_{ik}\delta_{jl} \delta(\br-\br')\delta(t-t').
\eqnend
The   particle    evolves   according   the    Second   Newton's   Law
(\ref{eq:partVel}) and receives the force  exerted by  the fluid
$\bl$  and  eventually  some  other  external (or  inter-particle)
potential  force $\bF(\bq,t)$.  In  turn, the  fluid phase receives back
from the particle a local source of  momentum density given by $-\S\bl$
(see Eq.  \ref{eq:consvMom}). This form guarantees the Third  Newton's Law
both globally  and locally  (see \cite{Usabiaga2012b} and  below).  In
passing  we  note  that,   in  contrast  to  friction-based  couplings
\cite{Giupponi2007,Duenweg2009}, we do  not assume any functional form
for the  fluid force $\bl$.  Instead,  $\bl$ is treated  as a Lagrangian
multiplier  to  impose  (at any instant)  the  no-slip  constraint
(\ref{eq:noSlip}).  This  allows to recover  the correct hydrodynamics
under  quite different flow  regimes; even  at large  Reynolds numbers
where  the drag  force has  a  strong convective  origin and  strongly
deviates       from       the       Stokes      (friction)       value
\cite{Usabiaga2013,pep-rafa}.

The appearance of $m_e$ in the particle equation of motion (\ref{eq:partVel}) 
reflects the Archimedes Principle, which states that the inertial mass of an object immersed in a fluid is equal to
its excess of mass $m_e$ over the fluid it displaces  $\rho_0 \vol$.
The nominal particle mass is then,
\eqn
\label{eq:particleMass}
m_p = m_e+ \rho_0 \vol.
\eqnend
Thus, for $m_e=0$ the particle is neutrally-buoyant and just follows the inertia
of the local fluid parcel. The particle kernel contains a fluid mass $m_f = \J\rho \vol$ 
whose equilibrium fluctuations are studied in Sec. \ref{densfluc}.

\subsection{Momentum conservation}
The total momentum in the particle kernel is then $\vol \J\bp = \J\left[\left(m_e +\rho \vol\right) \bv\right]$
which, using the no-slip constraint Eq. (\ref{eq:noSlip}), gives a kernel momentum density 
$\J \bp= m_e \bu/\vol + \J\bg$. The total momentum density field of the system (fluid and particle) is
just \cite{Usabiaga2012b,pep-rafa} $\bp(\br) = m_e \S(\bq-\br) \bu + \bg(r)$.
To better understand the coupled dynamics it is illustrative 
to write out the equations of motion for $\bp$ and $\J\bp$. 

Eliminating  $\bl$ from Eq. \ref{eq:partVel} and 
after some algebra with Eqs. (\ref{eq:consvDens})-(\ref{eq:noSlip}) one finds,
\begin{equation}
\label{eq:totmom}
\frac{\partial \bp}{\partial t} = -\nabla\cdot\left[\bP +\bg \bv + m_e \S \bu\bu\right] + \S\bF,
\end{equation}
which, for vanishing external force $\bF=0$, shows that rate change of
total momentum $\bp$ can be written in a conservative form. Therefore, $\bp$ is {\em locally} conserved
and obviously $\int \bp d^3$ is a constant of motion.

Taking averages in Eq. (\ref{eq:totmom}) and noting that
the material derivative concomitant to the particle is
\eqn
\fr{\dd \J \bg}{\dd t} = \J \left[
\fr{\partial \bg}{\partial t} + \bna \cdot (\bu \bg),
\right]
\eqnend
one gets,
\eqn
\label{kerdy}
\fr{\dd \J \bp}{\dd t} = -\J \bna \cdot \left[\bP + (\bv-\bu)\bg \right] + \bF/\vol.
\eqnend

The change rate of the kernel momentum  $d\J\bp/\dd t= m_e \dot \bu + \dd \J \bg/\dd t$
is driven by the local fluid pressure force 
$-\J\nabla\cdot \bP$ and by convective forces, proportional to the relative acceleration between the
particle and the fluid inside the kernel. The particle equation of motion can be also written as,
\eqn
\fr{m_e}{\vol} \dot \bu + \J \left[\frac{\partial{\bg}}{\partial t} \right] = - \J \nabla\cdot \left [\bP + \bg \bv \right] + \bF/\vol. 
\eqnend
The right hand side contains all the (driving and damping) 
forces arising in the acoustophoretic phenomena. As explained below, this term includes
two very different time scales. The radiation force builds up in the (fast) sonic time scale,
but the slow dynamics of the particle is driven by a balance between the time-averaged 
sonic force and friction.

\subsection{Energy conservation}
It  has   been  demonstrated  \cite{Usabiaga2012b}   that  the  no-slip
constraint  $\J\bv=\bu$ does  not insert  energy into  the  system.  A
necessary condition  for this result  is the adjoint  relation between
$\J$ and  $\S$ (Eq. \ref{adj}). It  is not difficult to  show that the
modified pressure field $\pi(\rho)$ does not introduces energy either.
The total energy field per unit  mass can be written as $e(\br) =v^2/2
+ \epsilon$ where the field $\epsilon$ is the specific internal energy
$\epsilon= \epsilon_0 - \pi/\rho$.  We do not consider exchange of heat
in  this work  and the  energy  $\epsilon_0$, of  entropic origin,  is
constant.   The differential  form  of the  First  Law is  then $d \epsilon  =
(\pi/\rho^2) d\rho$ and only includes  the reversible work done by the
pressure field $\pi$. The rate of total energy production can be shown
to be (see e.g. \cite{Groot-Mazur1984,pep-rafa}),
\begin{equation}
  \frac{d}{dt}\int \rho e d r^3= \int \rho \frac{d\,e }{dt} d r^3 =
 -\int \nabla \cdot (\bP \cdot \bv) d^3r + \int {\bf f}^{ext} \cdot \bv d^3 r.
\end{equation}
Using the Gauss integral theorem $\int  \nabla \cdot (\bP \cdot \bv) d^3r
=  \oint \bv  \bP \cdot  {\bf  n} d^3r$  (with ${\bf  n}$ the outwards surface
versor) hence, a way to introduce energy into the system consists on moving its boundaries
($\bv  \ne 0$ at  the boundary).   For an  ideal fluid (inviscid limit) $\bP  = (p +\S\Omega )\bs 1$,  
the input power equals the rate  of reversible work $-\oint p\,  \bv \cdot {\bf n}  d^2r $ on the system's  boundaries.  
It is noted that the total work done by the particle compressibility $-\oint  \S \bv \cdot {\bf n}  \Omega d^2r$ 
vanishes ($\S$ has compact  support). 
In a periodic  system the total surface  integral vanishes
identically and the only way to  introduce energy is to apply an external volume force ${\bf f}^{ext}$,
as explained in Sec. \ref{simulations}

\subsection{Equilibrium fluctuations}
\label{densfluc}

\begin{figure*}
\includegraphics[width=0.45 \columnwidth]{JrhoVsOmega306.eps}
\includegraphics[width=0.45 \columnwidth]{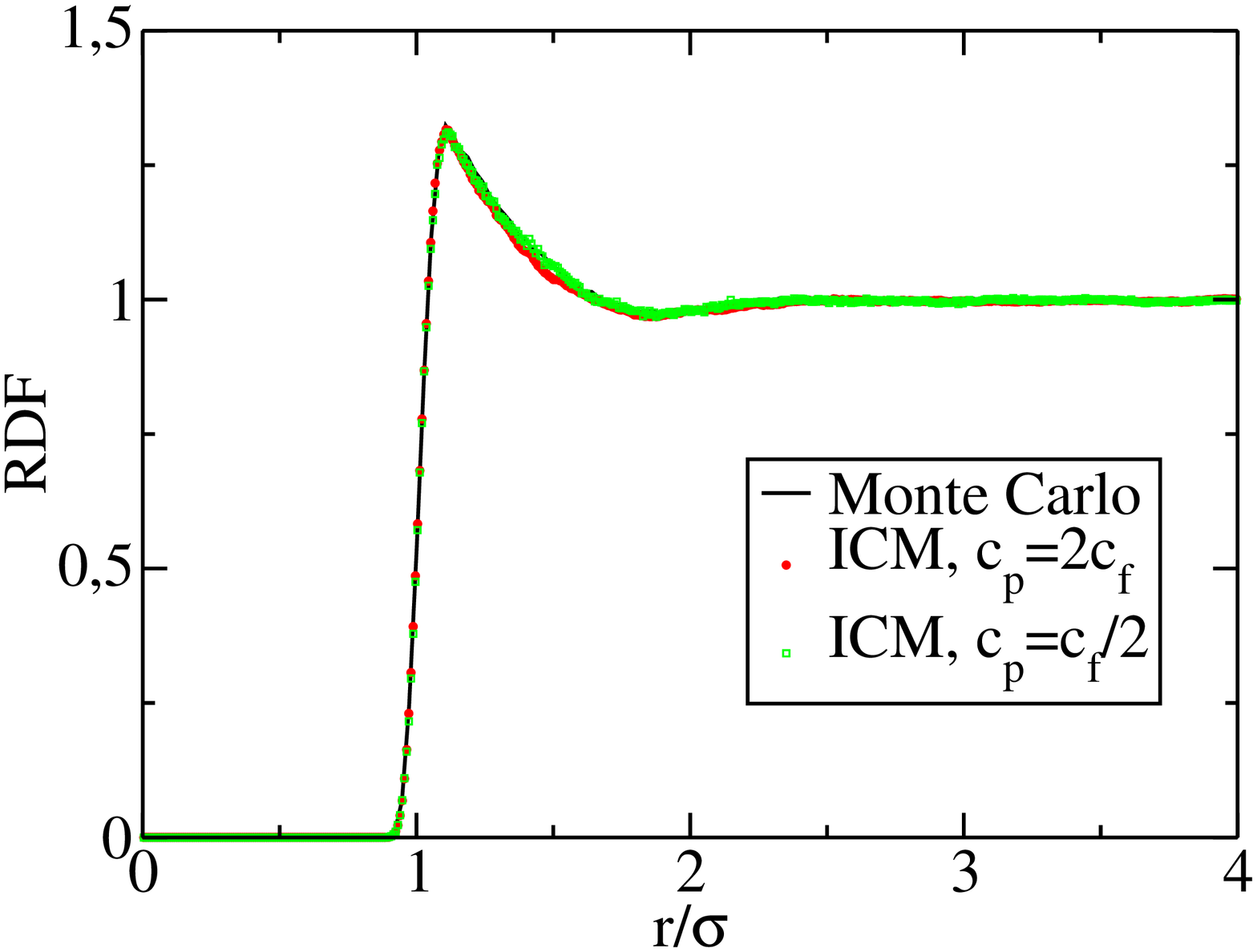}
\caption{Left, probability density function (PDF) for the 
average fluid density inside particle kernel, with varying particle compressibilities $\kappa_p=1/(\rho_0 c_P^2)$. 
Lines corresponds to the normal distribution with the grand canonical ensemble variance in Eq. (\ref{eq:densityPDF}). 
In all cases the fluid sound velocity is $c_f=4$ (parameters are given in table \ref{table:allSimulationParameters}).
Right, radial distribution of a set of particles at volume fraction $\phi=0.244$
interacting through a repulsive (truncated and shifted) Lennard-Jones
potential with strength $\epsilon=\kt$ with mass $m_e=0$. Comparison is made
between Monte-Carlo simulations and the hydrodynamic code with
two different particle compressibilities.}
\label{fig:densityFluctuations}
\end{figure*}

The  contribution $-\nabla (\S \Omega)$  to the  fluid momentum
equation is non-dissipative. The way to numerically verify this is
to show that the equipartition of energy remains unaltered upon adding the
particle compressibility term.  To do  so we evaluated the static structure
factor  of the longitudinal  velocity $S_{v,v}(q)$  in an  ensemble of
$N=1000$   compressible   particles   ($c_p=2c_f$)  interacting   with
repulsive  Lennard-Jones potential with strength $\epsilon=\kt$ 
and  volume fraction   $\phi=0.244$. As   expected,  the   structure   factor  is
$q-$independent $S_{v,v}(q) = \kt/\rho_0$, showing that the added
particle  compressibility   term  does  not   affect  the  fluctuation
dissipation  balance  \cite{Usabiaga2012b}.  Further  we  measured  the
radial  distribution  function  (RDF)  of  ``colloids''  with  different
compressibilities.  Results, in the right panel of Fig. \ref{fig:densityFluctuations} show that the RDF
is not essentially affected by the particle compressibility. This result is not however not
as  general as energy  equipartition. Acoustic  Casimir forces 
could, in  principle, alter the  structure of a  colloidal dispersion.
The thermo-acoustic Casimir forces are however small \cite{Bschorr1999}, although larger acoustic Casimir
forces can be triggered by {\em forced} white noise of strong amplitude \cite{Larraza1998}.

In the present approach the particle kernel can be sought as a small 
domain of fixed volume $\vol$ which encloses the particle 
and it is open to the fluid. As expressed in Eq. (\ref{jdp}), the particle compressibility is 
here translated as an excess in the isothermal compressibility of the fluid 
in the kernel. 

The mass of fluid in the kernel $m_f=\vol \J\rho$ fluctuates
and in equilibrium ($\langle \Omega \rangle=0$ and  $\langle \J\rho\rangle = \rho_0$)
its variance should coincide with the grand canonical ensemble prescription $\mathrm{Var}[m_f] = m_f \kt/c_P^2$.
The kernel-density variance should then be,
\eqn
\label{eq:densityVariance}
\mathrm{Var}\left[\left(\J\rho\right)^2\right] = \fr{\rho_0 \kt}{c^2_p \vol}.
\eqnend
In the weak fluctuation regime (assumed by the fluctuating hydrodynamics formulation \cite{Landau1987}) 
the density probability distribution should then be Gaussian,
\eqn
\label{eq:densityPDF}
P(\J\rho) = \pare{\frac{\vol c_p^2}{2 \pi \rho_{0} \kt}}^{1/2} \exp\pare{-\frac{\vol c_p^2}{2\rho_{0}\kt}(\J\rho-\rho_{0})^2}.
\eqnend
Figure \ref{fig:densityFluctuations} shows the numerical results obtained for $P(\J\rho)$ for particles 
with different compressibilities, immersed in a
fluid at thermal equilibrium. Results are compared with the 
grand-canonical distribution of Eq. \ref{eq:densityPDF}. 
We find excellent agreement, for particles with either larger or smaller
compressibility than the surrounding fluid (in Fig. \ref{fig:densityFluctuations} $c_f=4$, see Table I
for the rest of simulation parameters).  
As shown in Sec. \ref{numerics}, the variance of the kernel density can be used as a sensible measure of the 
convergence of the numerical scheme.

\section{Acoustic Forces}
\label{acousticforces}
A  central  application  of  the  present work  is  the  simuation  of
acoustophoresis of  small particles $(a>0.1  \mu\mathrm{m})$ suspended
in a  fluid subject  to MHz ultrasound waves.  Such process
which  is   receiving  renewed  attention  in  the   context  of  many
applications,   such  as   control  and   manipulation   of  particles
microfluidic devices.  We now  briefly explain its  essential features
and the reader is  refereed to Refs. \cite{Gorkov1962,Landau1987,Doinikov1997,Settnes2012} 
for a more comprehensive theoretical  description. 

We start by considering a fluid under otherwise quiescent  condition, which is submitted to an
oscillatory  mechanical   perturbation  (maybe  through   one  of  its
boundaries)  which creates a standing acoustic wave.  The
amplitude  of the  sound wave  is assumed  very small,  so  a standard
approach    \cite{Landau1987,Settnes2012}   consists   on    expanding   the
hydrodynamic fields  whose amplitude decrease as  increasing powers of
the wave amplitude. To second order,
\begin{eqnarray}
  \rho&=&\rho_0+\rho_1+\rho_2\\
  \bv&=& \bv_1+\bv_2.
\end{eqnarray}

The  time  dependence  of  any hydrodynamic  perturbative  field  (say
$\Phi_i$ with $i>0$) should  have a fast oscillatory contribution with
the same  frequency as the  forced sound wave, i.e.   $\Phi_i(\br,t) =
\phi(\br) \exp(i\omega t)$.  The average $\langle \Phi \rangle = (1/\tau) \int_0^{\tau} \Phi(\tau) d\tau$
over the wave  period $\tau = 2\pi/\omega$  vanishes.  Inserting this  expansion into the mass and momentum
fluid equations leads to a hierarchy of equations at each order in the
wave amplitude. At first order the set equations are linear so
the time-average of the first-order momentum change rate  yields no
resulting  mean  force.  However,  at  second order,  the  average  of
non-linear terms (such as $\langle \rho_0\bv_1 \bv_1\rangle$) do not vanish ($\langle \cos(\omega  t)^2\rangle =1/2$)  and  create the  so called  
radiation force.  The leading terms creating the radiation force
are already present in an inviscid fluid and for most applications 
viscous terms only lead to relatively small corrections \cite{Settnes2012}.
Viscous forces are only important near the particle surface  $r=a$, where
the oscillating fluid velocity field is enforced to match the particle
velocity.   At a distance  $\delta =  \sqrt{2 \nu/\omega}$ from the particle surface, called 
viscous   penetration  length or sonic boundary layer,  
the  fluid  inertia (transient term) $\rho  \partial_t \bv \sim \rho \omega  v$ becomes of
the  same order  than  viscous  forces $\eta  \nabla^2  \bv \sim  \eta
\delta^{-2} v$.  For $|r-a|>\delta$ the  fluid can be treated as ideal
(inviscid)  so the ratio $\delta/a$  determines  the  relevance of
viscous regime \cite{Settnes2012}. For large values $\delta/a \simeq 5 $ the 
acoustic force reach a plateau which corresponds to the transient (frictional) 
Stokes force \cite{Settnes2012}. Here we focus on  the inviscid  regime ($\delta << h$) where
we expect the inertial (instantaneous and energy  conserving) coupling will 
{\em  quantitatively} capture the acoustophoretic forces \cite{Usabiaga2013} 
on small particles with arbitrary acoustic contrast.

The force exerted by a standing wave on a spherical
particle was derived by Gor'kov for the case of an inviscid
fluid \cite{Gorkov1962} and recently extended to viscous
fluids by Settnes and Bruus \cite{Settnes2012}.
The primary acoustic force can be written in the form
\begin{equation}
\label{eq:acousticForce}
\bF_1 = -\bna U_{ac}   
\end{equation}
where the acoustic potential $U_{ac}$ is given in Eq. (\ref{eq:acousticPotential}).
For a sinusoidal wave along the $z$ axis with wavenumber $k$ the expression for
the force can be simplified to,
\eqn
\label{eq:acousticForce2}
\bF_1 = \fr{c_f^2 \Delta \rho^2 \vol k}{4 \rho_{0}} \pare{f_1 + \fr{3}{2}f_2} \sin(2kz),
\eqnend
In the inviscid fluid limit, the viscous layer $\delta=\sqrt{2\nu/\omega}$ is small
compared with the wave length $\lambda$ and the particles radius,
the coefficients $f_1$ and $f_2$ are \cite{Gorkov1962,Settnes2012},
\eqn
\label{f1}
f_1 &=& 1 - \fr{\kappa_p}{\kappa_{f}}= -\fr{\kappa_e}{\kappa_{f}} \\
\label{f2}
f_2 &=& \fr{2(\rho_p - \rho_{0})}{2\rho_p + \rho_{0}} = \fr{2 m_e}{2m_e + 3\rho_0 \vol}
\eqnend
where the particle density is $\rho_p = m_p/\vol = m_e/\vol + \rho_0$. 

In this work we extend the blob model to model a particle with finite compressibility $\kappa_p$.
Under the local pressure variations of an incoming sound wave
a compressible particle pulsates and in doing so it eject fluid mass in the form
of a spherical scattered wave. If the particle and fluid compressibilities do not match, 
the scattered fluid mass is ejected at a rate which differs 
from the flux of the incoming wave. This difference creates variations in the Archimedes force
which is expressed as a (monopolar) radiation force \cite{Crum1974,Settnes2012}.
The mass of fluid in the kernel is $m_f= \vol \J \rho$ so the mass 
ejected by pulsation of the particle volume, can be equivalently  expressed in terms of 
changes in the local fluid density.  Consider an incoming pressure wave $p_{in}$ which is scattered by the particle. 
The incoming density wave satisfies $\rho_{in}= \rho_0 \kappa_f p_{in}$, so if the particle were absent,
the mass of fluid in the kernel would be $\vol \J\rho_{in} =  \vol \rho_0 \kappa_f \J p_{in}$.
However, the particle modifies the local density according to
Eq. (\ref{kappao}) and the total mass inside the kernel is then 
$\vol \J\rho  = \J({\bs \kappa} p_{in}) \vol$  
with,
\begin{equation}
 \J \rho =  \left(1 + \frac{\kappa_e}{\kappa_f} \right) \J \rho_{in}.
\end{equation}
The scattered mass
\begin{equation}
 m_{sc}=  \vol \frac{\kappa_e}{\kappa_f} \J \rho_{in} 
\end{equation}
is then ejected at a rate,
\begin{equation}
\dot m_{sc}=  \vol \frac{\kappa_e}{\kappa_f} \frac{d}{dt} \J \rho_{in} = \vol \frac{\kappa_e}{\kappa_f} 
\left[\J (\partial_t \rho_{in})  + \J(\nabla \cdot \rho_{in} \bu )\right],
\end{equation}
where the prefactor $f_1=-\kappa_e/\kappa_f =1-\kappa_p/\kappa_f$ 
is in agreement with Gor'kov theoretical result \cite{Gorkov1962,Settnes2012}.
It is noted that the advective term $\J(\nabla \cdot \rho_{in} \bu)$ is a second order quantity 
neglected in theoretical analyses \cite{Settnes2012} for low Reynolds numbers,
however particle-advective terms need to be included in studies of larger bubbles 
at non-vanishing Reynolds \cite{Pelekasis2004,Garbin2009}.

\section{Acoustic forces: simulations}
\label{simulations}

To create a standing wave in a periodic box we employ a simple method that resembles the experimental setups \cite{Haake2005}.
We include a periodic pressure perturbation in all the cells at the plane
with coordinate $z=z_0$. The pressure perturbation has the form,
\eqn
\label{pext}
p^{ext}(t) = \Delta p_0 \sin(ck_0t) h \delta(z-z_0)
\eqnend
where $k_0=2\pi/L$ is the smallest wave number that fits into the simulation box
of length $L$. In the discrete setting the delta function should be understood 
as a Kronecker delta $h \delta(z-z_0) = \delta^K_{z z_0}$ so only the
cells at the plane $z=z_0$ are forced.

A solution for the density modes can be analytically obtained by 
inserting the forcing pressure (\ref{pext})  into the linearized 
Navier-Stokes equations and transforming the problem into the Fourier space. This leads to,

\eqn
\rho_k = \Delta \rho_{k} \sin(ck_0t+\phi) = \fr{k^2 \Delta p_0}{\sqrt{4\Gamma^2k^4(ck_0)^2 + ((ck)^2 - (ck_0)^2)^2}} \sin(ck_0t+\phi).
\eqnend
Where $\Gamma=\nu_L/2$ is the sound absorption coefficient (which, in absence of heat diffusion, equals half
of the longitudinal viscosity).
The singular pressure perturbation $\delta(z-z_0)$ excites 
all the spatial modes of the box. However, since $c \gg \Gamma k$,
the resonant mode $k=k_0$ is by far the 
dominant one and it is safe to assume that the incoming wave is just
a standing wave with wavenumber $k_0$,
\eqn
\rho_{in}(z,t)=  \Delta \rho_{k=k_0} \cos(k_0z)\sin(ck_0t+\phi)
\eqnend
The validity of this approximation requires working in the linear regime $\Delta \rho_0 \ll \rho_0$
(i.e. low Mach number) which is also satisfied in experiments.

We checked the validity of the present model against 
the theoretical expression for the (primary) radiation force in Eq. (\ref{eq:acousticForce2}),
by measuring the acoustic force felt by particles with 
different mass $m_e\ne 0$ or compressibility $\kappa_e \ne 0$ than the carrier fluid.
To measure the acoustic force at a given location, particles were bounded to an harmonic
potential $U_{\mathrm{spring}} = -(1/2) k_{\mathrm{spring}} \left(z-z_{eq}\right)^2$
with a given spring constant and equilibrium position $z_{eq}$.
The acoustic force displaces the equilibrium position
of the spring to an amount $\Delta l$ and its average gives the local 
acoustic force $F(\hat{z})= k_{\mathrm{spring}} \langle \Delta l \rangle$
where $\hat{z}=z_{eq}+\langle \Delta l\rangle$.
In order to conserve the total linear momentum of the system, we place two particles
at equal but opposite distances from the pressure perturbation plane $z=z_0$ (a wave antinode).
In this way the momentum introduced by each harmonic force cancels exactly.
Moreover to minimize the effect of secondary forces, particles were placed at different
positions in the $x,y$ plane. In most simulations the particles positions were at
${\bf r}_1=(1/2,1/2,3/8) L$ and ${\bf r}_2=(0,0,-3/8) L$.

\subsection{Monopolar acoustic forces}
\begin{figure*}[h]
\includegraphics[width=0.45 \columnwidth]{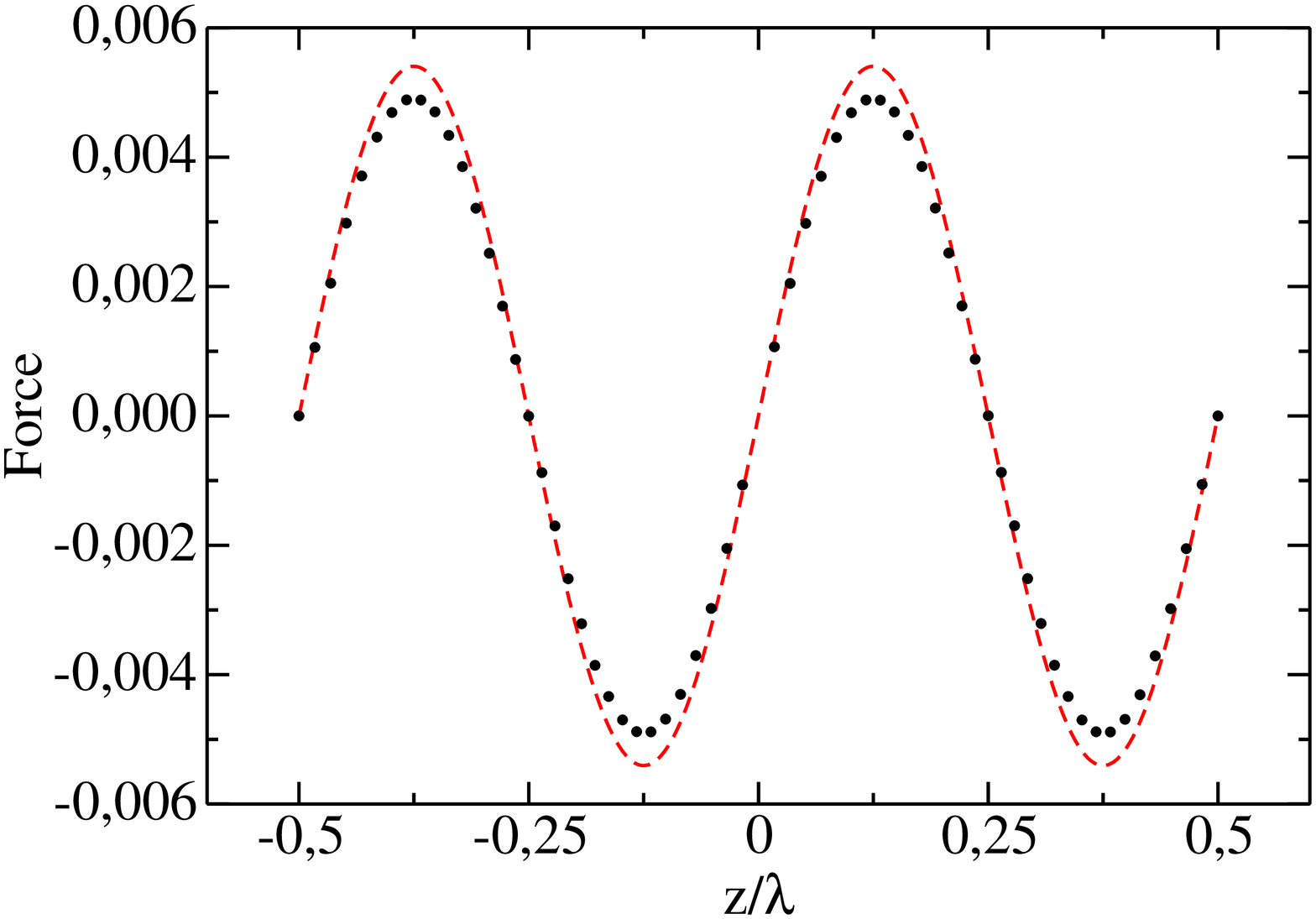}
\includegraphics[width=0.45 \columnwidth]{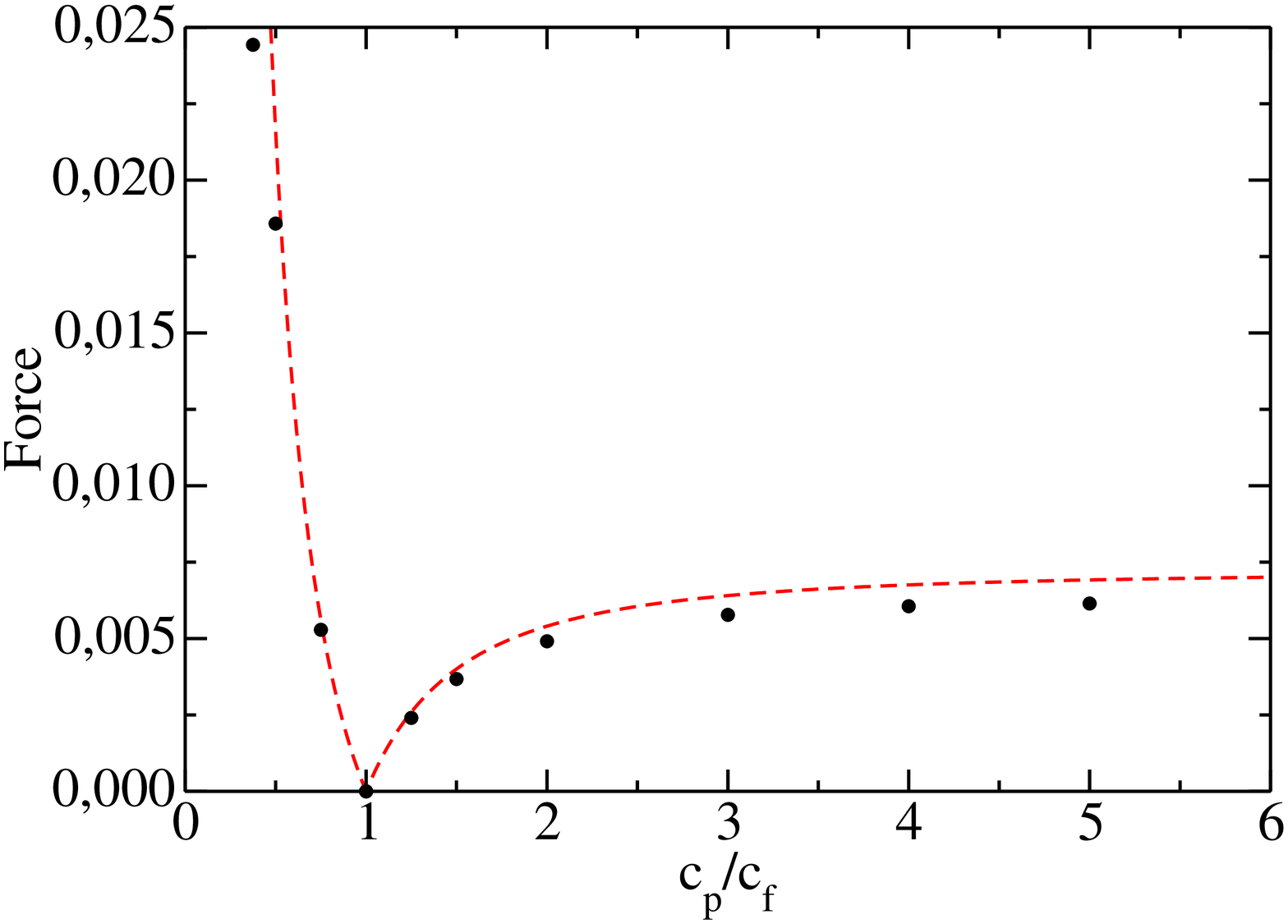}
\caption{
Left panel, acoustic force along the pressure wave for a neutrally-buoyant
particle with $c_p=2c_f$.
Right panel, maximum acoustic force versus the ratio $c_p/c_f$.
All the simulation parameters are given in table \ref{table:allSimulationParameters}.
}
\label{fig:force3}
\end{figure*}

According to the acoustic potential in Eq. (\ref{eq:acousticPotential}), 
neutrally buoyant particles ($m_e=0$) can only feel monopole acoustic forces proportional
to the deficiency in particle compressibility $-\kappa_e$ with respect to the carrier fluid.
[see $f_1$ in Eq. (\ref{f1})]. The  left panel  of  figure \ref{fig:force3}  represents the  acoustic
force observed in numerical simulations at different positions in the  plane of the  standing wave $z$.
The  particle speed  of sound  is $c_p=2c_f$,  which corresponds  to a
particle   less  compressible  than   the  fluid   [$\kappa_e=  -(3/4)
  \kappa_f$,     see     Eq.      (\ref{ke})].      Simulations     of
Fig. \ref{fig:force3} were  performed in a cubic periodic  box of size
$L=32\,h$    (see   Table    I    for   the    rest   of    simulation
parameters). Numerical  results exactly recover the  dependence of the
radiation force  with $z$ given  by the theoretical expression  of the
primary radiation force in Eq. (\ref{eq:acousticForce2}). However, the
force amplitude presents deviations of  up to about 10 percent.  These
deviations tend to  zero as the box size  is increased, indicating the
presence  of hydrodynamic  finite size  effects which, as  explained in Sec. \ref{secondaryf},
scale like secondary  acoustic forces between particles \cite{Crum1974}.

The right panel of figure \ref{fig:force3}
shows the maximum  value of the acoustic force  for different particle
compressibilities   (here,  in   terms   of  the   ratio  $c_p/c_f   = (\kappa_f/\kappa_p)^{1/2}$). 
It is noted  that while the dipole  scattering coefficient is bounded  $f_2 \in
(-2,1)$,  the   monopole  scattering  coefficient  is   not  $f_1  \in
(-\infty,1)$:  for incompressible  particles  it goes  to $f_1=1$  but
diverges if particles are infinitely compressible $c_p/c_f \rightarrow
0$.  This explains why ultrasound is an outstanding tool to manipulate
bubbles \cite{Garbin2009}.   As shown in  Fig.  \ref{fig:force3}, the
present  method correctly  describes  the divergence  of the  acoustic
force  in  the  limit  of large  particle  compressibility,  $\kappa_p \rightarrow \infty$.

\subsection{Dipolar acoustic forces}

\begin{figure*}[h]
\includegraphics[width=0.45 \columnwidth]{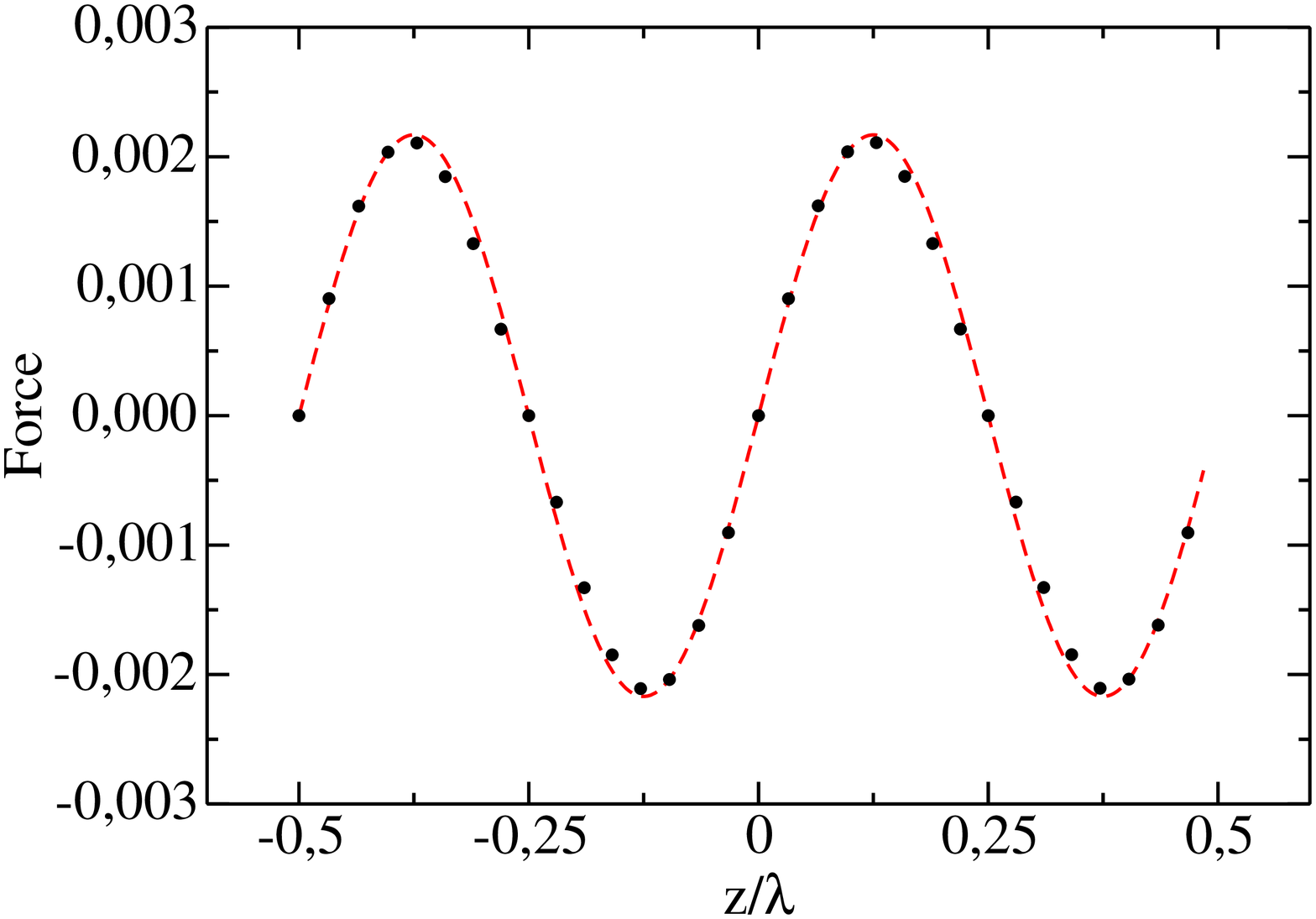}
\includegraphics[width=0.45 \columnwidth]{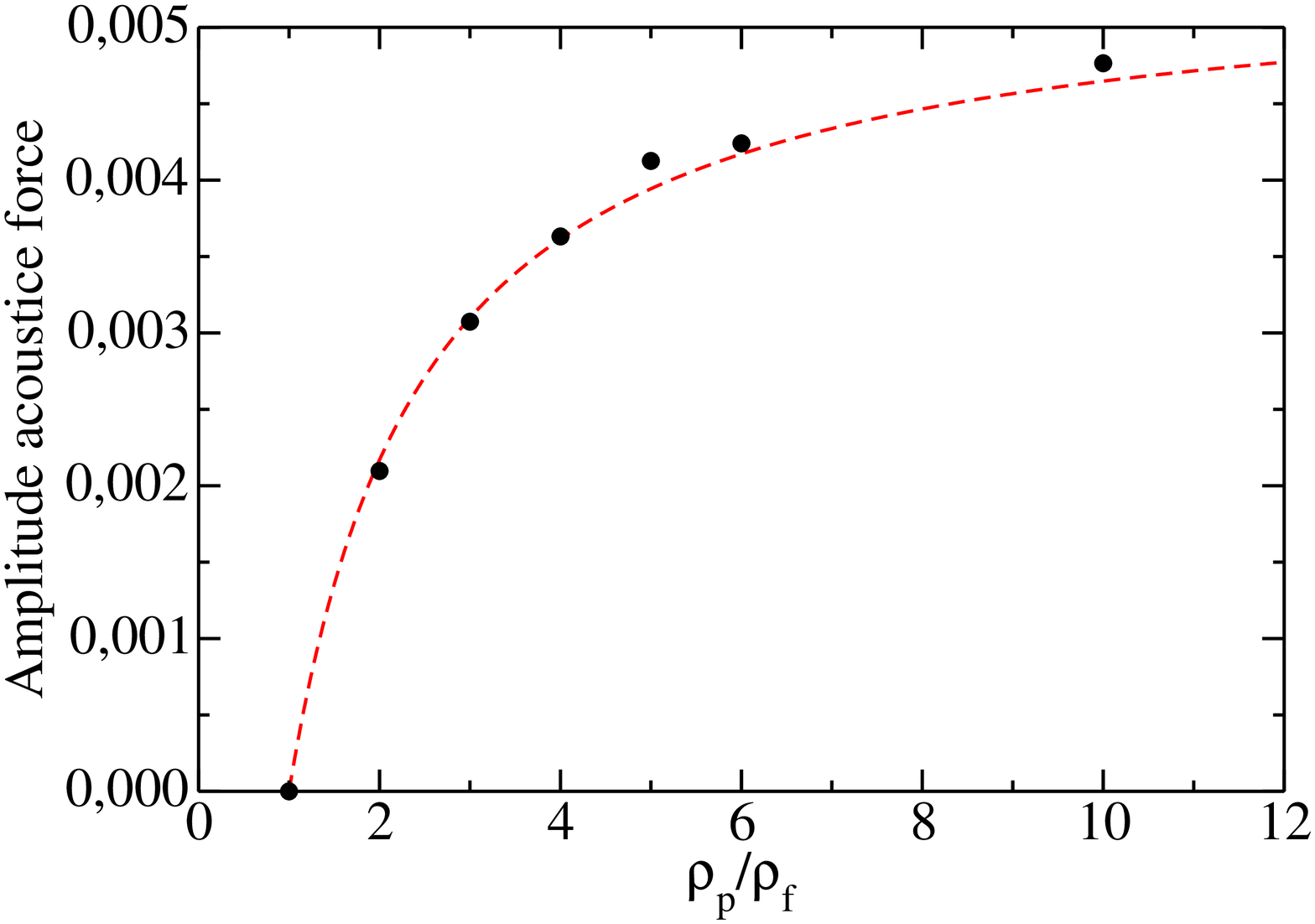}
\caption{Left panel, acoustic force along the pressure wave for
a non-neutrally buoyant particle with speed of sound $c_p=c_f$.
Right, maximum acoustic force versus the particle fluid density ratio
$\rho_p/\rho_0$ for $c_p=c_f$. Points represent the numerical results and
red lines the theoretical prediction.
All the simulation parameters are given in table \ref{table:allSimulationParameters}.
}
\label{fig:force}
\end{figure*}

In the left panel of figure \ref{fig:force} we plot
the acoustic force along the coordinate $z$ felt by
a particle with excess of mass $m_e=m_f$ and equal compressibility than the fluid 
$\kappa_e=0$. A perfect agreement is found
between the numerical results and Eq. \ref{eq:acousticForce2}.
In the right panel of the same figure
we show the dependence of the maximum acoustic force with the 
particle-fluid density ratio $\rho_p/\rho_0$.
Again, a quasi-perfect agreement ($1.5\%$ deviation) is observed when compared with
the theoretical expression for primary radiation force \ref{eq:acousticForce2}.

\subsection{Finite size effects: secondary radiation forces}
\label{secondaryf}

\begin{figure*}
\includegraphics[width=0.45 \columnwidth]{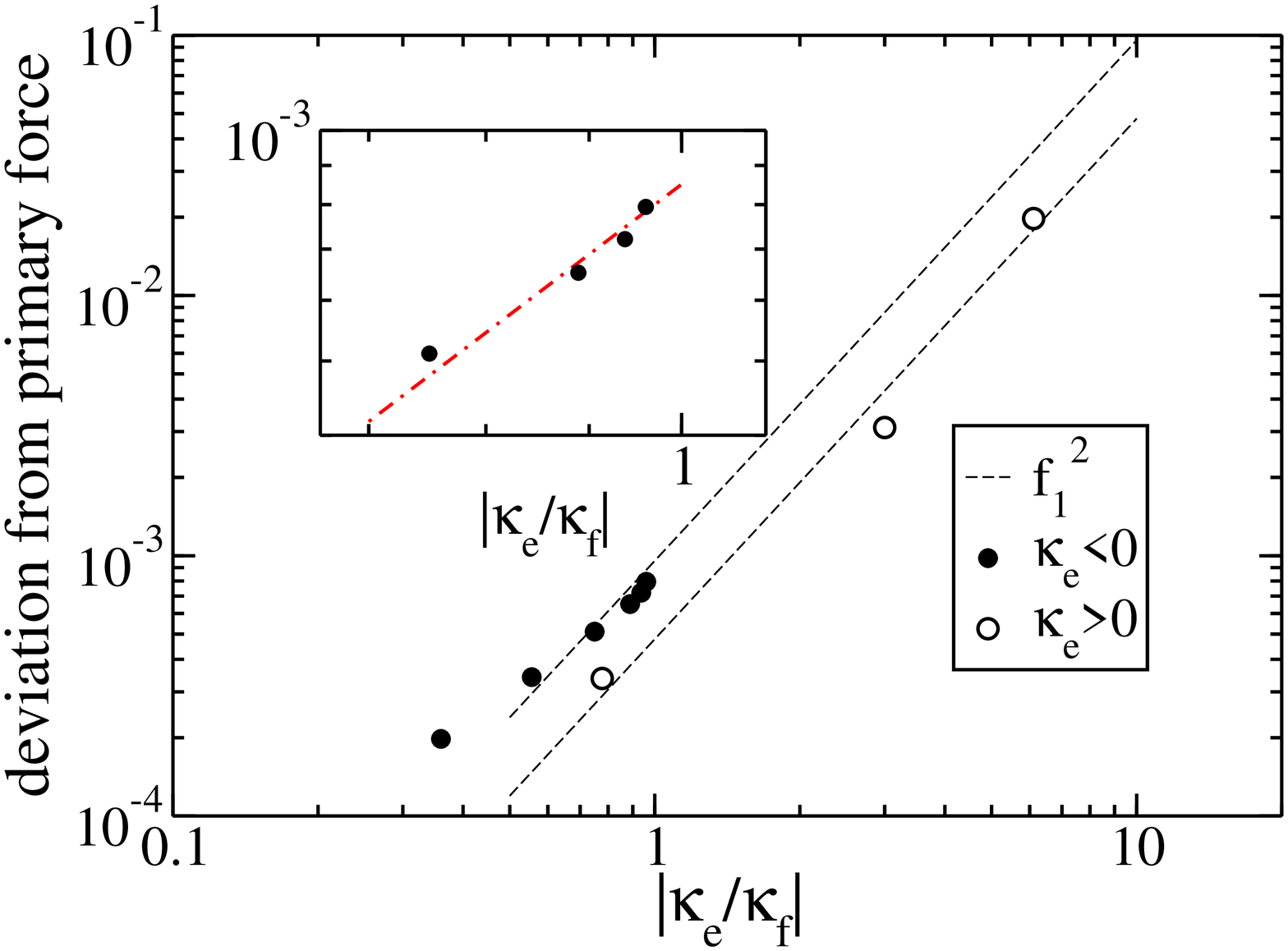}
\includegraphics[width=0.45 \columnwidth]{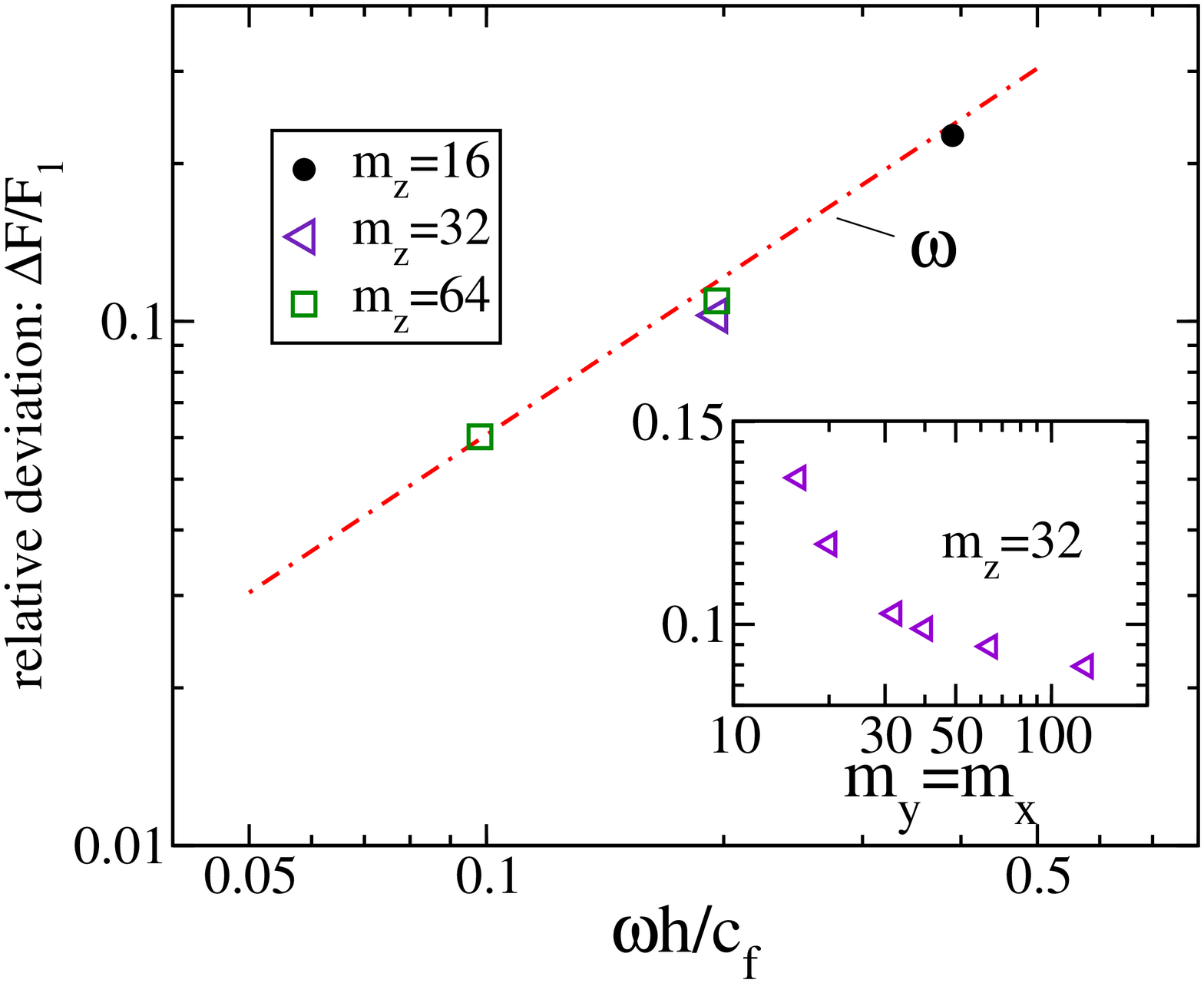}
\caption{
Left panel, deviation of the numerical force from the 
theoretical primary radiation force $\Delta F=F-F_1$ 
versus the ratio $\kappa_e/\kappa_f$ with $\kappa_e=\kappa_p-\kappa_f$. 
All parameters as in figure \ref{fig:force3}. 
Right panel, relative deviation $\Delta F/F_1$ versus the wave frequency $\omega$.
Main figure: simulations in cubic periodic boxes with $m_z$ cells per side; 
inset figure: rectangular boxes with $m_z=32$ and varying $m_x=m_y$
showing that the effect of the periodic images decreases with the system size.
}
\label{fig:errors}
\end{figure*}

To  understand  the   discrepancies  observed  between  numerical  and
theoretical expressions  for the primary radiation  force, we performed
simulations with different  box sizes $L$.  Results, in Fig.
\ref{fig:errors}, show that discrepancies between the numerical
and theoretical forces vanishes as $L$ increases and indicate that
these  deviations are not algorithmic or discretization  errors but
rather {\em finite size}  effects of hydrodynamic origin.  Notably, in
a periodic box, particles  can interact via secondary radiation forces \cite{Crum1974} 
arising from the  scattered  waves,  irradiated  by  each particle pulsation \cite{Garbin2009}. 
We now analyze the observed deviations to show that they have the signature of secondary radiation forces.

Secondary radiation forces, also called Bjerknes secondary forces,
depend on the particles' spatial configuration. The problem of elucidating the secondary forces 
from-and-to an array of scatters is certainly a difficult one \cite{Feuillade1995}, 
but approximate expressions have been proposed for a couple of interacting particles
at distance $d$, under certain conditions. In particular, for $R \ll d \ll \lambda$, 
Crum \cite{Crum1974}, Gr\"oschl \cite{Groeschl1998}  and others 
derived the following analytical expression for the secondary forces
for two particles at distance $d$ forming at angle $\theta$ with the incident wave is
\eqn
\label{sec_m}
\bF_2^{(p)} &=& -\frac{9}{4\pi} \vol^2 \ang{p^2_{in}(z)} \left[\frac{\omega^2 \rho_0 \kappa_e^2} {9 d^2} \right] {\bf e_r} \\
\label{sec_v}
\bF_2^{(v)} &=& \frac{3}{4\pi} \frac{m_e^2}{2\rho_0 d^4} \ang{v^2_{in}(z)} 
\left[(3 \cos^2 \theta - 1) {\bf e_r} + \sin(2 \theta) {\bf e_{\theta}}\right]
\eqnend
In general, however, the secondary forces depend on the phase difference between the 
field scattered from particle $1$  (at the particle $2$ location) and the vibration of particle $2$ \cite{Crum1974,Mettin1997}.
This phase relation is neglected in the derivation of Eqs. \ref{sec_m}, \ref{sec_v}, which
assumes that $p_{in}(z+d)=p_{in}(z)$ (same for the velocity field) and that both particle oscillates in phase.
Details of Bjerknes secondary forces are still under research\cite{Garbin2009,Pelekasis2004}, for instance,
in the case of bubbles, this phase difference might even lead to secondary force reversal (it is attractive for zero phase difference,
see Eq. \ref{sec_m}). 

Let us first analyze secondary forces resulting from an imbalance in the particle density
with respect the fluid density, $m_e\ne 0$ when particles have similar compressibility as the fluid $\kappa_e=0$.
In this case,  the scattered field has the form of a dipole and 
decays with the square of the distance \cite{Settnes2012}. Therefore,
secondary forces (dipole-dipole interaction) should decay as the 
fourth power of the distance, as expressed in Eq. (\ref{sec_v}). 
These type of secondary forces are thus short-ranged  (and small in magnitude)
so they do not induce finite size effects. Consistently,
we do not observe any trace of finite size effects in simulations on dipolar acoustic forces, as shown in Fig. \ref{fig:force}.

By contrast, particles with some excess in compressibility $\kappa_e \ne 0$ 
vibrate in response to the primary wave, acting as point-sources (monopoles) of fluid mass
and creating scattered density waves. These monopolar scattered fields decays like $1/r$ so
the secondary interaction between two particles decays with the square of their distance (see Eq. \ref{sec_m}). 
This means that secondary compressibility forces are long ranged and reach image particles beyond the primary box of the periodic cell.
Although the exact form of the multiple scattering problem leading to finite size effects in periodic boxes
is not easy to solve, it is possible to elucidate some of their essential features.
In our setup, due to symmetry, secondary forces are directed in $z$ direction (as the primary one)
so the total radiation force on one particle (say $i=1$) should be 
(summing up to pair reflections in the scattering problem), 
$F = F_1(z_i) + \sum_{j\ne i} F_2(r_{ij})$, with $j$ running over all particles (including periodic images)
and $F_1$ given by Eq. \ref{eq:acousticForce2}).

For any particle pair, the magnitude of $F_2$ is proportional to the product of the fluid mass ejected by 
each particle, i.e. to $\kappa_e^2$ (see Eq. \ref{sec_m}). Thus, for a given external wave amplitude $\Delta \rho$,
the difference between the force $F$ from simulations and the theoretical primary force $F_1$
should be proportional to,
\eqn
\label{fdif}
\Delta F \equiv F-F_1  \propto \left(\frac{\vol \omega}{c_f}\right)^2   \left(\frac{\kappa_e}{\kappa_f}\right)^2 
\eqnend
We have measured $\Delta F$ for several compressibilities ratios $|\kappa_e/\kappa_f|$ 
and frequencies $\omega$.  The left side panel of figure \ref{fig:errors} shows
$\Delta F$ against $|\kappa_e/\kappa_f|=|f_1|$
for a set of force measures with only differ in the value of $\kappa_e$.
As predicted by the scaling of secondary forces (\ref{fdif}), we get a quadratic dependence
$\Delta F \propto |\kappa_e/\kappa_f|^2$. A slight deviation from this
trend is observed for the smallest value of $|\kappa_e|$ considered
(see Fig. \ref{fig:errors}). Near $\kappa_e=0$ both forces (primary $F_1$ and secondary force)
tend to zero (see Fig. \ref{fig:force3}) and the evaluation of $\Delta F$ becomes more prone to numerical errors.
Values of $\Delta F$ for $\kappa_e>0$ (more compressible particles) 
and $\kappa_e<0$ (less compressible) were found to differ in a factor 2; the reason might
come from some change in the phase difference of the interacting particles taking place at $f_1=0$.

In the right panel of figure \ref{fig:errors} we show the relative difference $\Delta F/F_1$ 
obtained in simulations at different forcing frequencies $\omega$.
The primary force scales linearly with $\omega$ (see Eq. \ref{eq:acousticForce2})
so, according to Eq. \ref{fdif}, the relative difference $\Delta F/F_1$ 
should also scale linearly in frequency, as observed in Fig. \ref{fig:errors} (left).
Although an analysis of the total effect of 
multiple scatterings of secondary forces (Edwald summation)
is beyond this work, the inset of this figure shows that
the effect of scattered waves from periodic images decreases with the system size.

To  further  check the  resolution  of  secondary  acoustic forces  we
performed  some  tests  with   two  neutrally  buoyant  particles  and
compressibilities $\kappa_p= \kappa_f/2$.  Particles were
located at ${\bf r}_1 =  (0,0,z_0)$ and ${\bf r}_2 = (d,0,z_0)$, where
$z_0$ is  the plane of the  pressure antinode where  the primary force
vanishes.  As predicted by the theory (see Eq. \ref{sec_m}) the radial
secondary  forces were  found to  be attractive.   At  close distances
$d=[2-3]\,h$,  we  found  them  to  be in  very  good  agreement  with
Eq.  (\ref{sec_m}) although, at  larger distances  we found  that they
decay significantly slower than $d^{-2}$, probably due to the effect of
secondary forces  coming from the  periodic images.  In any  case, for
most practical  colloidal applications the effect  of secondary forces
is  small  and quite  localized.  It  tends  to agglutinate  close  by
colloids to form small clusters, but only after the main primary force
collects them in the node-plane  of the sound wave. Simulations showed
that this {\em local} effect  of the secondary forces is captured by the present method.

\subsection{Boltzmann distribution and standing waves}
Most of the experimental works on acoustophoresis
employ particles with diameters above one micrometer
or at least close to that size.  The reason is that the
acoustic force decays strongly with the particles radius
and below diameters of one micrometer other forces become
equally important in the nano-particle dynamics.
As stated previously, one of these forces is the streaming force \cite{Nyborg1965},
whose nature and structure is more difficult to control \cite{Bruus2012}. 
Advances in miniaturized devices and in experimental 
techniques makes easy to guess that acoustophoresis
will be soon  extended to smaller scales (see the recent work \cite{Johansson2012}).
An intrinsic limitation for this miniaturization process comes however from 
thermal fluctuations which strongly affect the dynamics of nanoscopic particles. 
Here we study how thermal fluctuations disperse sonicated particles
around the minimum of the acoustic potential energy.

\begin{figure*}
\includegraphics[width=0.65 \columnwidth]{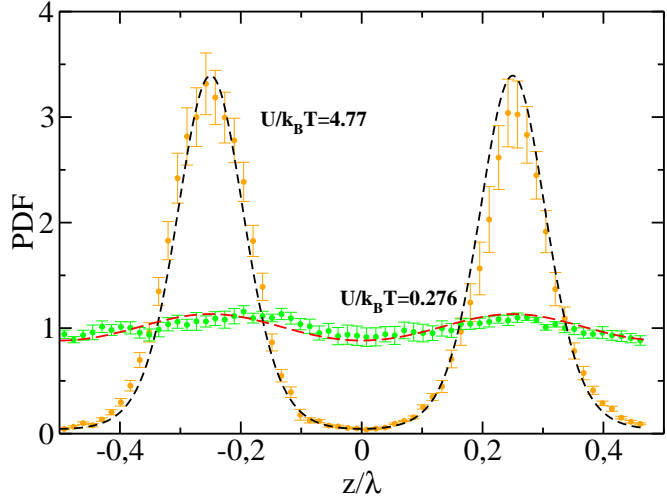}
\caption{PDF for particles inside an acoustic field.
Lines Boltzmann distribution. Circles PDF for a single particle
(volume concentration $2.4\cdot 10^{-4}$). 
Temperature $\kt=1$, the rest of the parameters are given in table \ref{table:allSimulationParameters}.}
\label{fig:pdfOneParticle}
\end{figure*}

A standing  waves exert  a first-order force  that oscillate  with the
same frequency  $\omega$ than  the primary wave  and averages  to zero
\cite{Bedeaux1974}.   Since the  diffusion  of the  particles is  much
slower than the wave period, the first-order force should not have any
effect in the slow (time-averaged)  dynamics of the particle, which is
driven by the  second-order radiation force.  If the  particle mass is
not  very large  (typical  particles-fluid density  ratio $m_p/m_f  \sim
O(1)$) the particle inertia, acting in times of $m_p/\xi = (2/9) (m_p/m_f) a^{2}/\nu$,  
is  also  negligible  in  the  time  scale  of  Brownian
(diffusive)  motion ($a^{2}/D$). This  indeed is  only true  provided a
large value of the Schmidt  number $Sc=\nu/D \sim \nu^2 \rho a/\kt>>1$
such  as  those  found  in  solid  colloid - liquid  dispersion  (here
$D=k_B T/(6\pi\rho\nu a)$  is  the  Stokes-Einstein  diffusion  coefficient).
Thus,  in  the  Brownian  time  scale, the  relevant  forces  are  the
radiation   force  $F_1$,  resulting   from  the   acoustic  potential
Eq. (\ref{eq:acousticPotential}), the  Stokes friction (which, assuming
$\langle {\bf  v}\rangle =0$,  is equal to  $6\pi\eta a {\bf  u}$) and
dispersion forces from fluid momentum fluctuations. Assuming there are
no  other  momentum  sources,  such  as secondary  forces  from  other
particles, and  that there  is no temperature rising from  conversion of
acoustic energy  into heat, the resulting time-averaged  motion can be
described by the Brownian dynamics  of a particle in an external field,
given  by  the  acoustic potential  (\ref{eq:acousticPotential}).  The
resulting  particle  spatial   distribution  should  then  follow  the
Gibbs-Boltzmann distribution, \eqn
\label{eq:particlePDF}
P(\br) \propto e^{-U_{ac}(\br)/\kt} \eqnend  This rationale was proposed in
an early work by  Higashitany {\em et al.} \cite{Higashitani1981}, who
found  a good  agreement  with experiments  in  very dilute  colloidal
suspensions.  For validation purposes,  the
simulations presented hereby are done within the range of validity of these 
approximations. Figure  \ref{fig:pdfOneParticle} shows the probability
density function  of the position of  a single particle  in a standing
wave, where different  wave amplitudes have been chosen  so as to vary
the depth of the acoustic potential well \ref{eq:acousticPotential}.
The  agreement between  the numerical  result and  the Gibbs-Boltzmann
distribution  is remarkably  good  and illustrates  the difficulty  in
collecting particles  as soon as dispersion  forces dominate, $U<\kt$.
The present method offers the possibility to investigate what happens if any of the above
approximations  fail; notably, in situations  where non-linear  couplings might
become relevant, such as the effect of colloidal aggregation, 
secondary forces between particles or advection by thermal velocity fluctuations \cite{Donev2011d}.

\begin{table}[h]
  \setlength{\tabcolsep}{12pt}
  \begin{tabular}{lcccc}
    \hline
    Figure & \ref{fig:densityFluctuations} & \ref{fig:force} & \ref{fig:force3} & \ref{fig:pdfOneParticle} \\
    \hline
    grid spacing $h$ & $10$ & $10$ & $10$ & $10$ \\
    number of cells & $32^3$ & $32^3$ & $32^3$ & $32^3$ \\
    fluid density & $1$ & $1$ & $1$ & $1$ \\
    shear viscosity $\eta$ & $0.5$ & $0.5$ & $0.5$ & $0.5$ \\
    bulk viscosity $\zeta$ & $0.5$ & $1$ & $0.5$ & $0.5$ \\
    fluid speed of sound $c_f$ & $4$ & $4$ & $4$ & $4$ \\
    wave frequency $\omega$ & - & $0.0784137$ & $0.0784134$ & $0.0784134$ \\
    pressure forcing $\Delta p$ & - & $0.005$ & $0.005$ & $0.005$ - $0.025$ \\
    density perturbation $\Delta \rho$ & - & $0.00240249$ & $0.00339202$ & $0.00339202$ - $0.0140955$ \\
    hydrodynamic radius $R_H$ & $9.9$ & $9.9$ & $9.9$ & $9.9$ \\
    particle's excess of mass $m_e$ & $0$ & $8000$ & $0$ & $0$ \\
    particle's speed of sound $c_p$ & $2$-$40$ & $4$ & $8$ & $8$ \\
    \hline
  \end{tabular}
  \caption{Variables and parameters of the particle (arbitrary units). 
The forcing frequency is $\omega =2\pi c_f/L$ with $L=m_z h$ the box size.
}
  \label{table:allSimulationParameters}
\end{table}

\section{Concluding remarks}
\label{conclusions}
This  work   presents  a   {\em  coarse-grained}  model   to  simulate
acoustophoretic phenomena on  small particles $O(\mu\mathrm{m})$.  The
model  is   based  on   an  Eulerian-Lagrangian  approach   where  the
(isothermal)  fluctuating  hydrodynamics  equations  are solved  in  a
staggered grid (finite volume scheme) and the colloidal particles move
freely in  space.  The communication between the  Eulerian lattice and
the  particle Lagrangian dynamics  is based  on the  Immersed Boundary
(IB) method however, here each  particle is described with a single IB
kernel.  The kernel  is  used  to i)  average  local fluid  properties
(e.g. velocity, density) and ii) to convert particle forces into 
a localized  force  density field,  which  acts  as  a source  of  fluid
momentum.  In this way, the particle-fluid interaction conserves local
momentum exactly. We use a  {\em kinematic} coupling between the fluid
and the particle which enforces that the kernel-average fluid velocity
($\J\bv$) equals the  particle velocity, $\J\bv=\bu$ \footnote{We have
  also tried other  {\em dynamic} couplings based on momentum $\J\bg =
  \rho_0 \bu$ and also $m_p \bu = \vol \J\bp$ 
  but did not observed any significant difference in the simulation
  results.  In fact ultrasound applications work at extremely low Mach
  numbers $\delta \rho << \rho_0$ and, in the zero Mach limit, all couplings coincide,
$\J\bg =\rho_0 \J \bv$}. The  essential property of this  type of coupling  is that it
instantaneously transfers momentum between  the particle and the fluid,
thus   resolving   the   inertia    of   both   particle   and   fluid
\cite{buscalioni_nic}. This instantaneous 
{\em inertial coupling}, as  we called it \cite{Usabiaga2013}, is required  to resolve ultrasound forces
which builds up in sonic  times $a/c$, several  orders of  magnitude faster
than friction $a^2/\nu$. 

The  second   novelty  of  the  present   method  is  the   use  of  a
minimal-resolution model  for the particles. We work  with the 3-point
kernel introduced by Roma and Peskin \cite{Roma1999} which only demand
27 fluid cells per particle.  Despite its computational efficiency and
simplicity the kernel is physically robust in the sense that it embeds
all  the essential  particle  properties (size,  mass  and, as  proved
hereby, compressibility).  Notably,  radiation forces on particles are
proportional to their volume, which in the present model is a constant
(position-independent) quantity $\vol  =8h^3$ pertaining to the kernel
shape and  mesh size  $h$.  In this  work the  kernel is also  used to
implement an  arbitrary  particle   compressibility by
embedding a  small domain with a different equation  of state \ref{jdp}.
Alternatively, the particle compressibility can be justified from a free energy functional constructed 
from the particle-fluid (potential) interaction. Here, such functional 
would have the form,
\begin{equation}
  {\cal F}\left[\rho,{\bq}\right]= \frac{\vol \epf}{2\rho_0} \left[\J(\rho-\rho_0)\right]^2,
\end{equation}
providing a local fluid chemical potential arising from the particle presence,
\begin{equation}
\mu= \frac{\delta {\cal F}}{\delta \rho} = \frac{\vol \epf}{\rho_0} \S(\bq-\br) \left(\J\rho-\rho_0\right).
\end{equation}
Any variation  in this  chemical potential would  then induce  a force
density  field $\rho  \nabla  \mu$ in  the  fluid.  A  Boussinesq-type
approximation, valid at low Mach number $\rho \nabla \mu \simeq \rho_0
\nabla \mu$ leads  to the present model equations.  In particular, the
fluid  momentum equation \ref{eq:consvMom}  can be  then written  in a
conservative  form $\nabla  \rho_0\mu =  \nabla \S  \Omega$  (see Eqs.
\ref{omega}  and  \ref{totp}).   A  rigorous  connection  between  our
blob-model  (based on  a mean  field approach)  and  a first-principle
derivation of the coupled fluid-particle equations is beyond the scope
of  the present  work.  We  believe however  that  such connection  is
possible  and  will  provide  clues  to the  interaction  free  energy
functional  which   ultimately,  stems  from   molecular  interactions
\cite{pep-rafa}.   This would certainly  open many  other applications
(wettability) to  the present mean field approach.   Here however, our
main  target  problem  is to  model  the  fluid  mass ejected  by  the
pulsation of  a colloid's  volume forced by  an ultrasound  wave.  The
main benefit  of Eq.  (\ref{eq:pressure})  is that it  translates this
difficult ``mechanical'' constraint at  the particle surface in a much
more  simple  ``thermodynamic''  language:  it just  becomes  a  local
density  change.   The  excellent  agreement between  simulations  and
theory  \cite{Landau1987,Gorkov1962,Settnes2012}   confirms  that  this
``translation'' works.

The present  approach can be  safely used to resolve  micron particles
under  several  MHz,  using  for  instance,  water  as  carrier  fluid
$\nu\simeq  10^{-6}\mathrm{m}^2/\mathrm{s}$.   It  is also  suited  to
sub-micron particles $O(0.1 \mu  \mathrm{m})$, where the thermal drift
\cite{Higashitani1981}  becomes significant and  one needs  to include
hydrodynamic  fluctuations (here  they  are treated  according to  the
Landau-Lifshitz formalism).   Methods for the  acoustophoretic control
of sub-micron  particles are now  appearing and indeed  require larger
frequencies  (up   to  40MHz  range)   \cite{Johansson2012}.   Another
potential  problem  in controlling  submicron  particles  is the  drag
created by  the streaming velocity (the  second-order average velocity
field  $\langle  \bv_2  \rangle$)   which  at  these  scales,  becomes
comparable  to the  radiation force  \cite{Bruus2012}.   The streaming
field $\langle v_2\rangle$ spreads over the acoustic boundary layer of
any  obstacle  (e.g.  walls)  creating,  by  continuity,  an array  of
vortices.   Streaming  can be  certainly  resolved  using the  present
scheme (see  Ref.\cite{Usabiaga2012} for a  description on how  to add
boundaries  in  the  fluctuating  hydrodynamic solver)  although,  for
validation purposes here we use periodic boxes ($\langle \bv_2 \rangle
=0$) and avoid this effect.

As  in  any  {\em   coarse-grained}  description,  the  present  model
introduces  some artifacts  which has  to be  taken into  account when
analyzing  simulation  results.  In  particular,  acoustic forces  are
proportional to the particle  volume $\vol$ which, in principle, could
be  used   to  define  a   particle  {\em  acoustic  radius}   $R_a  =
(6/\pi)^{1/3} h  \approx 1.2407 h$.  This  ``acoustic radius'' however
is {\em not}  the particle hydrodynamic radius, which  for the present
surface-less,  soft-particle  model  takes  a somewhat  smaller  value
$R_H=0.91\,h$  \cite{Usabiaga2013}.  The blob hydrodynamic radius
is calibrated using the Stokes drag on a sphere with no-slip surface \cite{Maxey1983}
(i.e. $g\pi\eta R v_0$ and we use $g=6$ to calibrate $R=R_H$).
The slow particle dynamics arises from the balance of 
the acoustic force and the Stokes drag and in practice, to match experimental particle trajectories one should
consider that the blob model has a slightly smaller effective skin friction
(i.e., $R=R_a$ yields $g=4.4$).

The present  model cannot  properly resolve  viscous effects
related to the acoustic boundary layer $\delta = \sqrt{2 \nu/\omega}$.
The radius $R$ of the  present one-kernel-particle model is similar to
mesh size $h$, so $\delta \sim R  \sim h$ and the flow
inside  the   viscous  layer is ill-resolved.   We  observe a
limited sensitivity of the resolved  dipolar forces 
to the size of the acoustic  layer. For instance,  the primary  force in
Fig. \ref{fig:force} corresponds to  $\delta \simeq 0.28 \,R_a$ and it
is  found to  be about  $2\%$ larger  than the  inviscid  limit result
($\delta \rightarrow 0$), however Settnes and Bruus \cite{Settnes2012}
predict  that viscous  effects  should increase  this  force in  about
$10\%$.

Nevertheless, the  present approach offers  a route to  describe these
finer details  by adding more computational resources  to the particle
description   (larger   object    resolution,   in   the   spirit   of
fluid-structure interaction  \cite{Peskin2002}).  We believe  the save
in   computational   cost  would   be   still   large  compared   with
fully-Eulerian (particle remeshing) schemes and would allow to resolve
the  acoustic  boundary layer  (around  ``arbitrary''  3D objects)  and
provide more  accurate descriptions  of secondary acoustic  forces and
multiple scattering interaction in multiparticle flows.

Comparison  with   theoretical  expressions  show   that  the  present
generalization of  the IC method accurate resolves  acoustic forces in
particles  with arbitrary  acoustic contrast  (any excess  in particle
compressibility and/or mass).   The benefit of this minimally-resolved
particle model is that although  it has a very low computational cost,
it   naturally   includes   the   relevant   non-linear   hydrodynamic
interactions between particles:  mutual hydrodynamic friction, history
forces  \cite{Garbin2009}  convective  effects and  secondary  forces.
Interesting non-trivial effects such as changes in  the wave
pattern  due  to multiple  scattering  \cite{Feuillade1995} or
sound absorption  by colloids  or  bubbles \cite{Kinsler_book,Riese1999}
can also be simulated \footnote{In this later problem, the rate of momentum 
dissipation inside a droplet or a bubble can be
also generalized  by embedding a local particle viscosity  inside the
kernel}. The code \cite{Usabiaga} has been written in CUDA and efficiently
runs  in  Graphical  Processor  Units  (GPU): we  have  verified  that
simulations  with  $O(10^4)$ particles  over  the colloidal  diffusive
scale are feasible in affordable computational times.

\section*{Acknowledgments}
We thank Aleksandar Donev, Pep Espa\~nol and Ignacio Pagonabarraga 
for fruitful discussions and suggestions to broaden the scope of this research. 
We are quite honored to acknowledge funding from the Spanish government FIS2010-22047-C0S and from the 
Comunidad de Madrid MODELICO-CM (S2009/ESP-1691). 

\begin{appendix}
\section{Numerical implementation}  
\label{numerics}

We present in this appendix the time-stepping to
solve the equations \ref{eq:consvDens}-\ref{eq:noSlip}.
The fluid and hybrid (fluid+particle) package (we call {\sc fluam}) have been coded in CUDA 
to run on Graphical Processor Units (GPU) and they can be downloaded
under GNU license \cite{Usabiaga}.
Detailed explanation of the  
numerical scheme for the fluid solver can be found elsewhere \cite{Usabiaga2012,Usabiaga2012b,Usabiaga2013}.
Here we focus on the fluid-particle interaction
and in particular in the pressure contribution made by
the particles.

\subsection{Spatial discretization}
The fluid solver, explained in detail in Ref. \cite{Usabiaga2012},
employs a staggered grid to solve the Navier-Stokes equations.
In this grid the scalar variables (i.e. density) are defined
at the cell centers, which are located at $\br_{\bs{i}}$.
On the other hand, vectors, like velocity or momentum,
are defined at the cell faces. For example, the x-component
of the velocity is defined at $\br_{\bs{i}}+\fr{h}{2} \hat{\bs{x}}$. 
This nature of the staggered grid should be taken into
account when interpolating or spreading variables.
Then, the averaging of the fluid density at the
particle position $\bq$ is given by
\eqn
\J \rho = \sum_{\bs{i}\in \mbox{\tiny{grid}}} h^3 \ker(\bq-\br_{\bs{i}}) \rho_{\bs{i}}
\eqnend
while the interpolation of the x-component of the velocity is
\eqn
\J v^x = \sum_{\bs{i}\in \mbox{\tiny{grid}}} h^3 \ker(\bq - (\br_{\bs{i}}+\fr{h}{2}\hat{\bs{x}})) 
v^x_{\bs{i}+\fr{h}{2}\hat{\bs{x}}}
\eqnend
The same precaution should be followed when spreading variables
at cell centers (i.e. pressure) or at cell faces (forces like $\bs{\lambda}$ or $\bs{F}$).

The kernel is defined as the tensor product of three
interpolating functions $\phi(r)$, one for each spatial direction $\alpha$
\eqn
\ker(\br) = h^{-3} \prod_{\alpha} \phi\pare{\fr{r_{\alpha}}{h}}
\eqnend
Although it is not necessary to factorize the kernel in this
form, this choice is easy to implement and it is known
to give good results \cite{Peskin2002,Duenweg2009}
even if the kernel $\ker(\bq)$ is no longer isotropic.
For the interpolating function $\phi(r)$ we employ 
the three points kernel of Roma and Peskin \cite{Roma1999} 
\eqn
\phi(r) = \left\{ \begin{array}{l l}
\fr{1}{3} \pare{1 + \sqrt{-3r^2 + 1}} & |r| \le 0.5 \\
\fr{1}{6}\pare{5-3|r|-\sqrt{-3(1-|r|)^2+1}} & 0.5 \le |r| \le 1.5 \\
0 & 1.5 < |r| 
\end{array}\right.
\eqnend
which has a good balance between 
its properties to hide the grid discretization to the particle
dynamics and its computational efficiency
(each
particle only interacts with $27$ cells in three dimensions)\cite{Peskin2002,Roma1999,Duenweg2009}.

\subsection{Temporal discretization}
Our temporal discretization is based on previous works
for deterministic incompressible flows \cite{Griffith2012a}
and it was presented in reference \cite{Usabiaga2012b}.
The scheme has the following substeps
\begin{enumerate}
\item Update the particle half time step
\eqn
\bq^{\n12} = \bq^n + \fr{\dt}{2} \J^n \bv^n
\eqnend
Note that the particle is advected by the fluid as it could have been expected
from the no-slip condition. In the averaging we employ the 
particle position at time $t^n=n\dt$ as indicated by the superscript $n$ on $\J$.
\item Calculate the external force acting on the particle
at time $t^{\n12} = (\n12)\dt$
\eqn
\bF^{\n12} = \bF(\bq^{\n12},t^{\n12})
\eqnend
\item Update the fluid state from time $t^n=n\dt$ to
time $t^{n+1}=(n+1)\dt$ to obtain the final density
$\rho^{n+1}$ and the \emph{unperturbed} velocity 
$\bvt^{n+1}$. 
During this substep we take into account the effect of the external force $\bF^{\n12}$
and the particle contribution to the pressure, but we do not impose
the no-slip condition, note the absence of the force $\bl$ on the equations
\eqn
\ps{t} \rho + \bna \cdot (\bg) &=& 0 \\
\ps{t} \bg + \bna \cdot (\bg\bv) &=& -\bna \pi(\rho,\bq^{\n12}) + \bna \cdot \bs{\sigma} + \S^{\n12} \bF^{\n12} 
\eqnend
To solve this set of equations we employ a third-order Runge-Kutta
scheme as explained shortly. During this substep the particle
is fixed $\bq^{\n12}=\mbox{const.}$ and so it is the external force $\bF^{\n12}$.
\item Calculate the impulse exchange between fluid and particle during the
time step
\eqn
\Delta \bg = \dt (\bl + \bF^{\n12}) = \fr{m_e m_f}{m_e + m_f} \pare{\J^{\n12} \bvt^{n+1} - \bu^n}
\eqnend
Where $m_f$ is the fluid mass dragged by the particle $m_f=\vol \J^{\n12} \rho^{n+1}$.
\item Update the particle velocity
\eqn
\bu^{n+1} = \bu^n + \fr{\Delta \bg}{m_e} = \bu^n + \fr{m_f}{m_e+m_f} \pare{\J^{\n12}\bvt^{n+1} - \bu^n}
\eqnend
\item Update the fluid velocity in a momentum conserving manner
\eqn
\bv^{n+1} = \bvt^{n+1} - \fr{\vol}{m_f} \S^{\n12} \Delta \bg = \bvt^{n+1} + \vol \S^{\n12}\pare{\bu^{n+1} - \J^{\n12}\bvt^{n+1}}
\eqnend
Note that a neutrally-buoyant particle ($m_e=0$) is simply advected
by the fluid $\bu^{n+1}=\J^{\n12}\bvt^{n+1}$, as it usually assumed in the IB method \cite{Peskin2002}. 
At the end of this substep the no-slip condition is satisfied in the form
$\bu^{n+1}=\J^{\n12}\bv^{n+1}$ for either neutrally or non-neutrally buoyant particles.

\item Conclude the time step by updating the particle position to time $t^{n+1}=(n+1)\dt$
\eqn
\bq^{n+1} = \bq^n + \fr{\dt}{2} \J^{\n12} \pare{ \bv^{n+1} + \bv^n}
\eqnend
\end{enumerate}
The scheme is second order for $m_e=0$
(provided that in the third substep the fluid state
is updated to at least second order accuracy). However, for $m_e\ne 0$ the scheme
is only first order although it shows a good accuracy.

In principle, any compressible solver can be used in the substep $3$, 
we employ the strong stability preserving, third-order accuracy,
explicit Runge-Kutta scheme \cite{Usabiaga2012,Donev2010b}
The scheme is based on a conservative discretization of the
Navier-Stokes equation of the form
\eqn
\ps{t} \bs{U} = - \bna \cdot \bs{F}(\bs{U},\bq,\bW,t)
\eqnend
where $\bs{U}=(\rho,\bg)$ is an array that collects the fluid
variables density and momentum and $\bs{F}(\bs{U,\bq,\bW,t})$ represent the
flux of the fluctuating Navier-Stokes equations. 
The flux depends on the particle position through the pressure field
$\pi(\rho,\bq)$ and also on the random numbers $\bW$ through the
stochastic fluxes.
The Runge-Kutta scheme 
consist on three substeps where it
calculates predictions at times
$t^{n+1}=(n+1)\dt$, $t^{\n12}=(\n12)\dt$ and the
final prediction at time $t^{n+1}=(n+1)\dt$.
In each substep
the following increment is calculated
\eqn
\Delta \bs{U}(\bs{U},\bq,\bW,t) = -\dt \bna \cdot \bF(\bs{U},\bq,\bW,t)
\eqnend
and the Runge-Kutta substep are
\eqn
\wtil{\bs{U}}^{n+1} &=& \bs{U}^n + \Delta \bs{U}(\bs{U}^n,\bq^{\n12},\bW^n_1,t^n) \\
\bs{U}^{n+1/2} &=& \fr{3}{4} \bs{U}^n + \fr{1}{4} \pare{\wtil{\bs{U}}^{n+1} + 
\Delta \bs{U}(\wtil{\bs{U}}^{n+1},\bq^{\n12},\bW_2^n,t^{n+1})} \\
\bs{U}^{n+1} &=& \fr{1}{3} \bs{U}^n + \fr{2}{3} \pare{\bs{U}^{n+1/2} + \Delta \bs{U}(\bs{U}^{n+1/2},\bq^{\n12},\bW_3^n, t^{\n12})}
\eqnend
The last substep can be written in the well known form
\eqn
\bs{U}^{n+1} = \bs{U}^n + \fr{1}{6}\pare{\Delta \bs{U}^n + 4\Delta \bs{U}^{n+1/2} + \Delta \wtil{\bs{U}}^{n+1}}
\eqnend
that shows that it is a centered scheme.
The combination of random numbers is such that guarantees a third-order
weak accuracy in the linear setting \cite{Donev2010b,Usabiaga2012b,Delong2013} and
they are
\eqn
\bW_1^n &=& \bW_A^n - \sqrt{3} \bW_B^n \\
\bW_2^n &=& \bW_A^n + \sqrt{3} \bW_B^n \\
\bW_3^n &=& \bW_A^n \\
\ang{\bW_C^n(\br_i) \bW_D^m(\br_j)} &=& \delta_{CD} \delta_{nm} \delta_{ij}
\eqnend

The only difference with previous
works is that here the pressure depends on the particle
position, which along the Runge-Kutta step is fixed $\bq^{\n12}=\mbox{constant}$. 
The three pressures used in the fluid update are
\eqn
\pi^n &=& c_f^2 \rho^n + \epf \vol \S^{\n12}\pare{\J^{\n12} \rho^n - \rho_{0}} \\
\wtil{\pi}^{n+1} &=& c_f^2 \wtil{\rho}^{n+1} + \epf \vol \S^{\n12}\pare{\J^{\n12} \wtil{\rho}^{n+1} - \rho_{0}} \\
\pi^{\n12} &=& c_f^2 \rho^{\n12} + \epf \vol \S^{\n12}\pare{\J^{\n12} \rho^{\n12} - \rho_{0}} \\
\eqnend

\subsection{Convergence analysis: comment on the variance of the kernel density}
\begin{figure}
\includegraphics[width=0.45 \columnwidth]{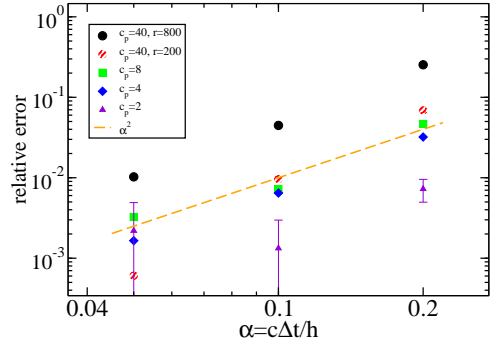}
\caption{Deviation between the input particle speed of sound $c_p=\sqrt{c_f^2+\epf}$ 
and that obtained from the best fit of the kernel density $\J\rho$ PDF to
the grand-canonical Gaussian distribution of Eq. (\ref{eq:densityPDF}) (see Fig. \ref{fig:densityFluctuations}).
The abscissa correspond to the CFL number $\alpha=c\Delta t/h$ where the speed of sound $c$ is the
maximum value between the fluid and particle speed of sound.
In the simulation with cell Reynolds number $r=200$ the viscosities where $\eta=\zeta=2$,
the rest of simulation parameters are given in Table I.}
\label{errorcp}
\end{figure}

We found that the PDF of  the interpolated density $\J \rho$ follows a
Gaussian  distribution for  all  the considered  cases.  However,  its
variance  presents some numerical  deviation if  large time  steps are
used. As  we said in  section \ref{ICM} this  variance can be  used to
measure the convergence order  of our scheme.  In figure \ref{errorcp}
we  present the  relative error  between the  input particle  speed of
sound $c_p=\sqrt{c_f^2+\epf}$ and  the numerical measure obtained from
the  variance  $\mbox{Var}[(\J\rho)^2]  = \rho_0\kt  \vol^{-1}/c_p^2$.
For neutrally  buoyant particles the scheme is  second order accurate,
as we anticipated. It is interesting  to note that when the cell Reynolds
number $r=ch/\nu$ is large, the errors are larger for a given speed of
sound and  time step. The  cell Reynolds number measures  the relative
importance of the advection relative to the viscous terms and it seems
that high advective terms reduce the accuracy of the present scheme.

\end{appendix}

\bibliography{biblio}

\end{document}

%% file: myNewCommands.tex
\newcommand{\eqn}{\begin{eqnarray}}
\newcommand{\eqnend}{\end{eqnarray}}
\newcommand{\kt}{k_B T}
\newcommand{\bs}[1]{\boldsymbol{#1}}
\newcommand{\ps}[1]{\partial_{#1}}
\newcommand{\pare}[1]{\left( #1 \right) }
\newcommand{\fr}[2]{\frac{#1}{#2}}
\newcommand{\wtil}[1]{\widetilde{#1}}
\newcommand{\mc}[1]{\mathcal{#1}}
\newcommand{\ang}[1]{\langle #1 \rangle}

\def\epf{\epsilon_{pf}}              
\def\ker{\theta_h}              

\def\vol{\mathbb{V}}              
\def\bp{{\bs p}}              %
\def\bP{{\bs P}}              %
\def\bu{{\bs u}}              %
\def\bv{{\bs v}}              %
\def\br{{\bs r}}              %
\def\bg{{\bs g}}              %
\def\bn{{\bs n}}              %
\def\bF{{\bs F}}              %
\def\dd{\mathrm{d}}              %

\def\bna{\bs \nabla}
\def\II{\bs I}

\def\n12{n+\fr{1}{2}}
\def\dt{\Delta t}
\def\S{\bs{S}}

\def\bvt{\bs{\tilde{v}}}

\def\bW{\bs{W}}
\def\bq{\bs{q}}

\def\J{\bs{J}}
\def\bl{\bs{\lambda}}